\documentclass[prb,aps,twocolumn,superscriptaddress,10pt]{revtex4-2}

\usepackage{graphicx}% Include figure files
\usepackage{bm}% bold math
\usepackage{hyperref}% add hypertext capabilities
\usepackage[mathlines]{lineno}% Enable numbering of text and display math
\usepackage[utf8]{inputenc}
\usepackage{setspace}
\usepackage[T1]{fontenc}
\usepackage{physics}
\usepackage{chemformula}
\usepackage{amsfonts}
\usepackage{amsmath}
\usepackage{xcolor}
\usepackage[shortlabels]{enumitem}
\usepackage{makecell}
\usepackage{float} 

\usepackage[scr = dutchcal]{mathalfa}
\usepackage{dsfont}

\usepackage[export]{adjustbox}

\newcommand{\editor}[2]{%
  \expandafter\newcommand\csname #1note\endcsname[1]{%
    \textcolor{#2}{(\textbf{#1:} \textit{##1})}}%
  \expandafter\newcommand\csname #1\endcsname[1]{%
    \textcolor{#2}{##1}}%
  \expandafter\newcommand\csname #1cancel\endcsname[1]{%
    \textcolor{#2}{\sout{##1}}}%
  \expandafter\newcommand\csname #1change\endcsname[2]{%
    \textcolor{#2}{\sout{##1} ##2}}%
  \newenvironment{#1text}{\color{#2}}{\color{black}}
}
\definecolor{Blu}{rgb}{0.00,0.00,1.00}
\definecolor{Red}{rgb}{1.00,0.00,0.00}
\definecolor{Orange}{rgb}{1.00,0.45,0.00}
\definecolor{Green}{rgb}{0.360,0.6,0.36}
\definecolor{tangerine}{rgb}{0.944,0.522,0}
\definecolor{brown}{rgb}{0.633,0.156,0.156}
\definecolor{lime}{rgb}{0.5,1.0,0.0313}
\definecolor{limedark}{rgb}{0.333, 0.666, 0.020}
\definecolor{applegreen}{rgb}{0.55, 0.71, 0.0}
\definecolor{green1}{rgb}{0.0, 0.5, 0.0}
\definecolor{green2}{rgb}{0.25, 0.5, 0.016}
\definecolor{BluBondi}{rgb}{0.00,0.58,0.71}
\definecolor{Cyan}{rgb}{0.00,0.30,0.30}

\editor{AG}{tangerine}
\editor{FM}{limedark}
\editor{FMau}{Red}
\editor{GC}{BluBondi}

\hypersetup{
	colorlinks = true,
	allcolors= blue,
}

\newcommand{\w}{\omega}

\newcommand{\p}{^\prime}

\newcommand{\pp}{^{\prime\prime} }

\newcommand{\pert}{^{\scriptscriptstyle{(1)}} }
\newcommand{\unpert}{^{\scriptscriptstyle{(0)}} }

\DeclareMathAlphabet{\mathsfit}{\encodingdefault}{\sfdefault}{m}{sl}
\SetMathAlphabet{\mathsfit}{bold}{\encodingdefault}{\sfdefault}{bx}{sl}

\newcommand{\myoverset}[2]{%
  \mathrel{\vbox{\offinterlineskip\ialign{%
    \hfil##\hfil\cr
    $\scriptstyle {#1}$\cr
    \noalign{\kern-.1ex}
    ${#2}$\cr
}}}}

\newcommand{\xc}{\mathrm{xc}}

\newcommand{\Hxc}{\mathrm{Hxc}}
\newcommand{\loc}{\mathrm{loc}}

\newcommand{\scf}{\text{scf}}
\newcommand{\ext}{\text{ext}}
\newcommand{\inveps}{\epsilon^{-1}}

\newcommand{\ZD}{\mathrm{2D}}

\usepackage[final]{changes}
\makeatletter
\@namedef{Changes@AuthorColor}{blue}
\colorlet{Changes@Color}{blue}
\makeatother

\begin{document}
\preprint{APS/123-QED}

%\title{Variational formulation of dynamical electronic response functions in Bethe-Salpeter, (screened) Hartree-Fock, Hybrid-DFT approaches}
\title{{Variational formulation of dynamical electronic response functions in presence of nonlocal exchange interactions}}
\author{Giovanni Caldarelli}
 \email{giovanni.caldarelli@uniroma1.it}
\affiliation{Dipartimento di Fisica, Università di Roma La Sapienza, Piazzale Aldo Moro 5, I-00185 Rome, Italy}
\author{Alberto Guandalini}
\affiliation{Dipartimento di Fisica, Università di Roma La Sapienza, Piazzale Aldo Moro 5, I-00185 Rome, Italy}
\author{Francesco Macheda}
\affiliation{Dipartimento di Fisica, Università di Roma La Sapienza, Piazzale Aldo Moro 5, I-00185 Rome, Italy}
\author{Francesco Mauri}
\affiliation{Dipartimento di Fisica, Università di Roma La Sapienza, Piazzale Aldo Moro 5, I-00185 Rome, Italy}

\begin{abstract}
{We consider the dynamical electronic response function in theoretical frameworks that include nonlocal exchange interactions, such as the Bethe-Salpeter equation with the frequency independent approximation of the screened interaction, Hartree-Fock, and range-separated Hybrid DFT approaches.
Within these pictures, we demonstrate that any time-dependent electronic linear response function allows for a formulation which is variational in the electronic density matrix.}
To achieve our goal, we consider the usual form of a response function, written in terms of a screened and a bare electronic vertices (`bare-screen'), and perform an exact rewriting in terms of purely screened electronic vertices (`screen-screen'). Within the `screen-screen' formulation, the response function can be written as a stationary point of a functional of the exact density matrix. Further, we show that the imaginary part of any electronic response can be written in the form of a generalized Fermi Golden Rule, by introducing an exact complementary rewriting in terms of vertices related by complex conjugation (`screen*-screen'). 
The screen-screen formulation can be further extended partitioning the electronic interaction in separate contributions, expressing the response in terms of partially screened electronic vertices (`partial screen-partial screen'), preserving the stationary properties.
We numerically validate the effectiveness of our formalism by calculating the optical conductivity of graphene, which exhibits strong excitonic effects. {To do so, we solve the Bethe-Salpeter Equation on a tight-binding model, including exchange effects in the response of graphene.} Our findings show the advantages of the variationality of the  screen-screen formulation over the others both in convergence properties and robustness with density-matrix approximations.
\end{abstract}

\maketitle

\section{Introduction}
Following linear response theory \cite{mahan2013many,giuliani2005quantum}, the response functions can be written in terms of a dressed (or screened) vertex and a bare (or unscreened) one. We refer to this formulation as `bare-screen'. 

Nonetheless, if some approximations for the vertices are performed, the bare-screen structure yields results that are not the closest to the exact solution. Let us take the example of the phonon self-energy, which can be seen as a response function where the vertices are represented by the electron-phonon couplings. In this case, it has been shown in Ref.\ \cite{piscanec2004} for static perturbations (see Eq.\ (2) and footnote [21] of this paper), extended to time-dependent perturbations in Ref.\ \cite{calandra2010adiabatic} and recently confirmed in Ref.\ \cite{berges2023phonon}, that an exact rewriting of the response function in terms of dressed vertices corresponds to a variational formulation in terms of the electronic density. It is thus clear that any approximation of the self-consistent density, and hence of the self-consistent vertices, affects the value of the response function to the second order, contrary to the bare-screen case which is affected to the first order. We refer to this variational formulation as the `screen-screen' approach. This functional formulation is closely related to that used for the static dielectric tensors by Gonze et al.\ in Ref.\ \cite{PhysRevLett.68.3603} within the Kohn-Sham, time-independent density-functional-theory frameworks.

%WE EXTEND PREVIOUS DEMONSTRATIONS
In this work, we extend the results of the Refs.\ \cite{piscanec2004} and \cite{calandra2010adiabatic} for a generic response function, and for a wider class of problems. In particular, we consider systems that can be described via generalized Kohn-Sham equations \cite{Baer2018}, i.e.\ containing nonlocal electronic interactions. We operate neglecting memory effects in the screening of the electronic interactions, as per the adiabatic approximation of time-dependent density functional theory (TD-DFT)~\cite{Ullrich2011,Marques2012,rocca2008turbo}.
Several theoretical  schemes can be recovered as particular cases of our framework, such as time-dependent Hartree-Fock~\cite{McLachlan1964TDHF,Bernasconi2012TDHF} (TD-HF), TD-DFT~\cite{Ullrich2011,Marques2012} with (global~\cite{Becke1993Hyb,Perdew1996Hyb} or range separated~\cite{
Yoshinobu2008RangeSep1,Seth2012RangeSep2,Zhan_23} hybrid functionals), and the Bethe-Salpeter equation~\cite{Hedin_1965,Strinati_1988,Onida_2002} (BSE) within the static-screened exchange (SX) approximation~\cite{martin2016interacting}. Nonetheless, our results can be generalized including memory effects into the Hartree-exchange-correlation part as of the electronic interaction, placing our approach in a wider context.

%%%%% === stress importance of exchange in optics, and epc 
%%
Such theoretical models are not only crucial to reproduce the optical properties of materials, but also to correctly describe the electron-phonon interaction in several systems.
Indeed, the inclusion of explicit exchange effects is crucial to account for the excitonic features in optical spectra of insulators~\cite{Rohlfing_2000} and low-dimensional materials~\cite{Spartaru_2004,Yang2009,Ferreira_2019,Qiu_2013}.
Excitonic effects have proved to be relevant also for molecular systems~\cite{Jacquemin_2015,Blase_2020}, in particular to describe charge-transfer excitations~\cite{Blase_2011}.
%%% aggiungere referenze impatto di ottica 
Moreover, a large impact of the exchange interaction on the electron-phonon coupling  was observed in graphene \cite{Basko_2008,Venanzi_2023,Graziotto2024-zk,PhysRevB.109.075420}, in strongly-correlated and charge-density wave materials and in superconductors \cite{yin2013correlation,pamuk2016spin,calandra2015universal,hellgren2017critical,romanin2021dominant,li2019electron,li2021unmasking,li2024two}.

The central achievement of our work consists in obtaining an exact rewriting of the response function in terms of screened electronic vertices, that is variational with respect to the one-body electronic density matrix. Our result is in close analogy to the screen-screen approach of Ref.\ \cite{calandra2010adiabatic}, but valid for a wider class of electronic interactions and any linear response.
In fact, the generality of this response formalism enables the investigation of virtually any type of electronic response, including the interatomic force constants used to compute phonons, and the electronic polarizability used to obtain optical or finite momentum transfer properties.

We complement our study by presenting another formulation, referred to as `screen*-screen', first introduced in Ref.\ \cite{allen-book} for DFT-like formalisms. Here, the response function is rewritten exactly in terms of two vertices that are one the complex conjugate of the other. We show that the screen*-screen formulation is not variational in the density matrix, but it allows the rewriting of the imaginary part of the response function exactly as a generalized Fermi Golden Rule. Being positive definite, it can be used to justify semiclassical Boltzmann equation approaches, where the dissipation is viewed as a scattering probability.

%== partial ==% 
Moreover, assuming that the electronic interaction can be partitioned in two distinct contributions, we present a formulation of the response function in terms of partially screened vertices. We refer to this formulation as the `partial screen-partial screen response'.  In this scheme, part of the interaction is contained in the vertices and the remainder is included in the electron-hole propagator. We further demonstrate the variationality of the partial screen-partial screen response with respect to the one-body electronic density matrix. The partial screen-partial screen approach is particularly appealing for pratical cases e.g.\ in BSE approaches. In these cases, we show that separating the interaction in the Hartree and exchange part results in a partial screen-partial screen response variational with respect to the electronic density. This feature enables the evaluation of BSE corrections to the response function using the  e.g.\ the electronic density obtained in a static DFT linear response theory calculation \cite{giannozzi1991ab,gonze1997dynamical}. If the relevance of the BSE corrections to the electronic density are limited, the variationality of the partial screen-partial screen formulation keeps the resulting error on the response under control. Interestingly, an analogous partial-screen partial-screen formulation was recently proposed in the contest of the exact many-body response in Ref.\ \cite{stefanucci2024phononlinewidthsscreenedelectronphonon}.

 % is particularly appealing, if we want to devise a practical scheme to evaluate the  BSE corrections to the phonon self-energy on top to  a standard static phonon calculation, based on DFT linear response theory \cite{giannozzi1991ab}. In this case, we can leverage the fact that both the dynamical and BSE effects  produce small relative-corrections to the static DFT electronic densities $\rho^{(b/a)}_{\rm DFT}$ in the range of frequencies corresponding to the phonon scale.

%PRACTICAL METHODOLOGY
We implement a computational procedure based on a set of self-consistent equations for the electronic density matrix to determine the response functions. The level of our approximation is equivalent to the SX+BSE methods, as developed e.g.\ in Refs.~\cite{Rocca2010BseStern,Rocca2012BseStern}.
We illustrate our developments with an application to graphene. As anticipated, in graphene excitonic effects are relevant in the description of several spectroscopic features, such as the logarithmic divergence of the group velocity near Dirac point~\cite{siegel2011many,Guandalini_2024}, the $\pi{-}\pi^*$ plasmon peak features in optics~\cite{Yang2009,Yang2011} and low momentum-transfer electron-energy loss spectroscopy~\cite{guandalini_2023}. Also, excitonic effects are expected to be important to determine the steepening of the Kohn anomaly and the enhancement of the electron-phonon coupling for phonons at zone border \cite{Basko_2008,Venanzi_2023,Graziotto2024-zk,PhysRevB.109.075420}.

In this work, we focus on the optical conductivity in the various formulations for the response function, but we pose the foundations for the calculation of exchange effects in electron-phonon related properties. We numerically demonstrate that the screen-screen approach is variational with respect to the electronic density matrix, while the others are not. 

As further proof, we compute the optical conductivity at a generic frequency $\omega$ by using the self-consistent density matrix obtained at a fixed frequency $\omega_0$. For $\omega {\simeq} \omega_0$, we obtain that the screen-screen approach approximates the exact optical conductivity with an error that is quadratic in the frequency, while bare-screen and screen*-screen display a linear error. 

The paper is organized as follows: in Sec.\ \ref{sec: linear response} we introduce the generalized Kohn-Sham equations and the linear response formalism used throughout the manuscript. In Sec.\ \ref{sec:sb ss s*s}, we formulate the response functions in terms of the screened vertices. We introduce the bare-screen, the screen-screen and the screen*-screen response functions. We show that the screen-screen response function can be seen as a stationary point of a functional, and it is thus variational in the density matrix. We show that the screen*-screen formulation allows an exact rewriting of the imaginary part of the response as a generalized Fermi Golden Rule. 
In Sec.\ \ref{sec: pp} we consider a repartition of the electronic interaction in two separate terms. We formulate the response functions in terms of partially screened vertices and partially interacting electron-hole propagator. We introduce the partial screen-partial screen response function and show that it is variational in the density matrix.
In Sec.\ \ref{sec: retardation} we discuss how to generalize our results relaxing the adiabatic approximation and including memory effects into the Hartree-exchange-correlation part of the electronic interaction.
In Sec.\ \ref{sec: application to graphene}, we present the application of our formalism, where we demonstrate the variationality of the screen-screen formulation computing the optical conductivity of graphene. Finally, in Sec.\ \ref{sec: conclusions} we draw the conclusions of our work.

\section{Electronic linear response functions with nonlocal exchange interaction} \label{sec: linear response}

In this Section, we summarize the linear response formalism that we use to describe the response equations of an electronic system subject to an external perturbation where nonlocal exchange interactions are included. 

We consider a class of systems for which the single-particle orbitals $\{ |\psi_{i\sigma}\rangle \}$ subject to a external time-dependent potential $\hat V_{\ext}(t)$ evolve according to the generalized time-dependent Kohn-Sham equations  
\cite{bauernschmitt1996treatment,rocca2008turbo,Baer2018GenTDDFT}
\begin{align}\label{schrodinger exact}
         i\hbar \pdv{}{t} |\psi_{i\sigma}(t)\rangle &{=} \left[-\frac{\hbar^2 \grad^2}{2m} +\hat V_{\text{scf}}(t)\right]|\psi_{i\sigma}(t)\rangle,\\
         \hat{V}_{\text{scf}}(t)&= \hat{V}_{\Hxc}(t)  + \hat{V}_{\ext}(t) \label{V_scf = Vx + Vext}  .
\end{align} 
$\hat{V}_{\Hxc}(t)$ is the self-consistent Hartree+exchange-correlation potential;\ $i, \sigma$ indicate respectively the electronic and spin state. In this work, we only consider spin independent Hamiltonians, hence we drop the $\sigma$ spin subscript for brevity. In the following, we express the matrix elements of nonlocal operators in Schrodinger representation as
\begin{equation}
    \langle\bm r|\hat O(t)|\bm r'\rangle {=} O(\bm r, \bm r\p,t) .
\end{equation}
The generalized potential $\hat{V}_{\Hxc}(t)$  depends self-consistently on the electronic wavefunctions. It is useful to introduce the  one-body electron density matrix, defined as
\begin{align}
    \label{exact density}
    n(\bm r,\bm r\p, t) = \sum\limits_{i}f_{i} \psi_{i}(\bm r \p,t)^*\psi_{i}(\bm r, t)  .
\end{align}
where $\psi_{i}(\bm r, t)=\langle \bm r|\psi_{i}(t) \rangle$ and $f_{i}$ is the Fermi-Dirac occupation of the state $|\psi_{i}\rangle$.
We define the electronic density as twice the diagonal part of the density matrix
\begin{align}
\rho(\bm r, t)=2n(\bm r,\bm r, t),
\end{align}
where the factor $2$ follows from spin degeneracy. In analogy with common applications of TD-DFT \cite{bauernschmitt1996treatment,rocca2008turbo}, we retain the adiabatic approximation in which 
\begin{equation} \label{V Hxc adiabatic}
    \hat{V}_{\Hxc}(t) \simeq   \hat{V}_{\Hxc}[\hat{n}(t) ]  .
\end{equation}
In considering the time dependence of  $\hat V_{\Hxc}(t)$ only through the density matrix at the same time, we neglect memory effects in the electronic interaction \cite{burke2005time}.
This amounts in considering a frequency independent approximation of the interaction kernel.

At variance with the TD-DFT approach, we consider a generalized nonlocal Hamiltonian $\hat{V}_{\Hxc}[\hat{n}(t) ]$. We call `local' $\hat{V}_{\Hxc}^{\text{loc}}$ the part of the Hamiltonian that depends on the electronic density as in TD-DFT, while the remainder is referred to as `nonlocal', i.e.\
\begin{align}
 %   \begin{split}
    V_{\Hxc}(\bm r, \bm r\p, t) &=  V_{\Hxc}^{\text{loc}}(\bm r,t)\delta(\bm r{-}\bm r\p)+ V^{\text{nloc}}_{\text{xc}}(\bm r,\bm r\p,t) ,\label{V Hxc exact} \\
    V_{\Hxc}^{\text{loc}}(\bm r,t)&=\frac{\partial E_{\Hxc}[\rho]}{\partial \rho(\bm r)}\Big{|}_{\rho(\bm r)=\rho(\bm r,t)}.    \label{V Hxc = dE/drho}
   % \end{split}
\end{align}
In Eq.\ \eqref{V Hxc = dE/drho}, $E_{\Hxc}[\rho]$ is the static DFT functional in which, if hybrid functionals are used, the double-counting terms associated to the exact exchange are removed. The potential $V_{\Hxc}^{\loc}(\bm r,t)$ in Eq.\ \eqref{V Hxc exact} contains both the Hartree and the local part of the exchange-correlation potential. For the nonlocal part of the Hamiltonian, we select the following form 
\begin{equation}
    \label{Vx (r,r)}
    V^{\text{nloc}}_{\text{xc}}(\bm r,\bm r\p,t)= -n(\bm r,\bm r\p, t)W(\bm r,\bm r')  .
\end{equation}
{ Depending on the choice of $W(\bm r, \bm r\p)$ in Eq.\ \eqref{Vx (r,r)}, we encompass different approximations of the electronic interaction. If $W (\bm r, \bm r\p){=} v(\bm r{-}\bm r\p)$, where $v(\bm r{-}\bm r\p){=}e^2/(|\bm r{-}\bm r\p|)$ is the Coulomb interaction, we recover the exchange potential used in Hartree-Fock schemes.
Range-separated hybrid functionals~\cite{Zhan_23} are obtained through $W(\bm r,\bm r\p) {=} \alpha(\bm r{-}\bm r\p)v(\bm r{-}\bm r\p)$, where $\alpha(\bm r{-}\bm r\p)$ is a well suited function used to separate and tune the ratio between short and long range parts of the exchange~\cite{Yoshinobu2008RangeSep1,Seth2012RangeSep2}. Finally, the SX approximation is obtained by setting $W(\bm r,\bm r\p) {=} \int \dd{\bm r_1} \inveps(\bm r,\bm r_1) v(\bm r_1 {-} \bm r\p)$ where $\inveps$ is the static inverse dielectric function in the random-phase approximation (RPA). }
% For every choice of the approximation of the exchange Hamiltonian in Eq.~\eqref{Vx (r,r)}, the local potential  $V_{\Hxc}^{\text{loc}}(\bm r,t)$ must be chosen accordingly to avoid double counting.

From now on, integration over spatial variables with numbers as subscript is always implicit, i.e.\ we adopt the following convention:
\begin{equation}
    \label{implicit integration}
    f(\bm r,\bm r\p,\bm r_1,\bm r_2) g(\bm r_1, \bm r_2) {=} \!\!\int\!\!{\dd{\bm r_1}}{\dd{\bm r_2}}f(\bm r,\bm r\p,\bm r_1,\bm r_2) g(\bm r_1, \bm r_2),
\end{equation}
where the spatial integrals are performed over the whole volume of the system. 
We express the analytic continuation of Fourier transforms  of retarded quantities in complex frequency $z{=}\omega {+} i \eta$ as 
\begin{equation}
    \label{fourier transform z}
    f(z) = \int\limits^{\infty}_{-\infty}\dd{t} e^{izt} f(t).
\end{equation}
The expression is well defined for $\eta{>}0$ due to the causality condition of $f(t)$, i.e.\ $f(t){=}0$ for $t{<}0$. Physical observables are obtained in the limit $\eta{\to}0^+$, to be considered after the thermodynamical limit \cite{kubo2012statistical}.
 
\subsection{Linear response}
 We suppose that the external potential is differentiable with respect to two independent time-varying parameters $a(t), b(t)$, which contain all the time-dependence of the external potential. We consider the completely general expansion 
\begin{align} 
      &  \hat V_\ext[a(t),b(t)] = \hat V^{\text{nl}\,(a)}_\ext a(t) {+} \hat V^{\text{nl}\,(b)}_\ext b(t)  \nonumber\\
       & {+}{\frac{1}{2}}\mqty[a(t), b(t)] {\cdot} \mqty[\hat V_\ext^{\text{nl}\,(a,a)} &\hat V_\ext^{\text{nl}\,(a,b)} \\\hat V_\ext^{\text{nl}\,(a,b)} & \hat V_\ext^{\text{nl}\,(b,b)}] {\cdot} \mqty[a(t)\\ b(t)]+\ldots \,.\label{potentialexp}
\end{align}
%The expansion of Eq.\ \eqref{potentialexp} is completely general.
In particular, we consider the case of external operators that are non local (`nl') in Schrodinger representation, i.e.\ $[\hat{V}_\ext^{\text{nl}\,(a)},\hat{\bm r}]{\neq} 0$. In this framework, we can define four-points response functions. Further, we can include practical situations where the effective perturbation is nondiagonal in space, as in the case of the ionic motion in a nonlocal pseudopotential picture \cite{martin2020electronic}. 

In linear response, the density matrix induced by the presence of the perturbation that couples to $b(t)$ is 
\begin{align}
       2n(\bm r,\bm r\p,t) {=}\!\! \int\limits^{+\infty}_{-\infty}\!\!\dd{t}\p L(\bm r,\bm r\p,\bm r_1 ,\bm r_2,t\p) V_{\ext}^{\text{nl}\,(b)}(\bm r_1,\bm r_2) b(t{-}t\p)   ,
\end{align}
where the factor 2 takes into account spin degeneracy, and $L$ is the retarded interacting electron-hole propagator, i.e. $L(t){=}0$ for $t{<}0$.

We now indicate $n^{(b)}(\bm r,\bm r\p,t)$ as the retarded response to the specific perturbation $b(t){=}\delta(t)$. Its expression in frequency domain is
\begin{equation} \label{n = LV0}
   2n^{(b)}(\bm r,\bm r\p,z) = L(\bm r,\bm r\p,\bm r_1 ,\bm r_2,z) V_{\ext}^{\text{nl}\,(b)}(\bm r_1,\bm r_2)  .
\end{equation}
For a generic perturbation $b(t)$, the density response in frequency domain can be deduced as $n(\bm r,\bm r\p,z)= n^{(b)}(\bm r,\bm r\p,z) b(z)$, where $b(z)$ is the Fourier transform of $b(t)$.  {For the remainder of this paper, we will only deal with linear response. Therefore, we will use the term electronic density matrix to address the induced density matrix rather than the exact ground-state density matrix introduced in Eq.\ \eqref{exact density}. }

Expanding at linear order in the external field the self-consistent potential of Eq.\ \eqref{V Hxc exact} we get
\begin{align}
V_{\scf}^{(b)}(\bm r,\bm r\p, z) &= V^{\text{nl}\,(b)}_{\ext}(\bm r,\bm r\p) \nonumber \\
&{+}K(\bm r, \bm r\p, \bm r_1, \bm r_2) n^{(b)}(\bm r_1, \bm r_2,z). \label{V_scf(r) = V + Kn}
\end{align}
The nonlocal kernel of the interaction $K$ is defined as 
\begin{align}
K(\bm r, \bm r\p,\bm r\pp, \bm r{'''})=&\frac{\delta V_{\Hxc}(\bm r, \bm r')}{\delta n(\bm r{''}, \bm r{'''})} \label{dVhxc/dn}\\
=&2f_{\Hxc}(\bm r,\bm r'')\delta(\bm r{-}\bm r')\delta(\bm r''{-}\bm r''') \nonumber\\
&- W(\bm r,\bm r') \delta(\bm r{-}\bm r'') \delta(\bm r' {-} \bm r''')  ,\label{kerneleq}
\end{align}
where
\begin{equation}\label{fdef}
f_{\Hxc}(\bm r,\bm r'')=\frac{e^2}{|\bm r - \bm r''|}+\frac{\delta V_{\xc}^{\loc}(\bm r)}{\delta \rho(\bm r'')}
\end{equation}
is the adiabatic Hartree plus exchange-correlation (`Hxc') kernel.
Using the linear response theory for time-dependent generalized Kohn-Sham equations \cite{casida1995recent}, the linear order expansion of the density of Eq.\ \eqref{exact density} is instead written as (see Appendix \ref{sec-app: td linear response} for a derivation):
\begin{equation}\label{n(r) = L0V  (r)}
    2n^{(b)}(\bm r,\bm r\p,z) {=}  L^{0}(\bm r, \bm r\p, \bm r_1, \bm r_2,z) V_{\scf}^{(b)}(\bm r_1,\bm r_2, z). 
\end{equation}
In Eq.\ \eqref{n(r) = L0V  (r)}, $L^0$ is the bare electron-hole propagator, defined as
\begin{align}
    L^{0}(\bm r, \bm r', \bm r''\!, \bm r'''\!\!,z) = 2 &\sum_{ij}\frac{f_j\unpert{-} f_i\unpert}{\hbar z {-} (\varepsilon_i\unpert {-} \varepsilon_j\unpert)}\psi_i\unpert(\bm r)\psi_j\unpert(\bm r')^*\nonumber \\  &\times \psi_i\unpert(\bm r'')^*\psi_j\unpert(\bm r''') \label{L0(r)} ,
\end{align}
where the superscript ${}^{(0)}$  indicates quantities computed without external perturbations, and the factor 2 accounts for spin degeneracy. 

A diagrammatic representation of Eqs.\ \eqref{V_scf(r) = V + Kn}, \eqref{kerneleq}, and \eqref{n(r) = L0V  (r)} is given in Table \ref{table:diagrams}. In particular, the coupled expressions of Eqs.\ \eqref{V_scf(r) = V + Kn}, \eqref{n(r) = L0V  (r)} represent the generalization of the TD-DFT equations to the case of nonlocal kernel \cite{runge1984density,Rocca2010BseStern,Rocca2012BseStern} . They can be solved with an iterative method. The process starts from the `unscreened' density matrix $n_0^{(b)} {=} L^{0} V_\ext^{\text{nl}\,(b)}$ and updates the potential $V_\scf^{(b)}$ through the application of the interaction kernel  $K$, as for the Sternheimer equations~\cite{baroni2001phonons}, until self consistency is achieved. 

It is also interesting to express the interacting electron-hole propagator $L$ in terms of the bare one, $L^{\rm (0)}$.
Using Eqs.~\eqref{V_scf(r) = V + Kn} and \eqref{n(r) = L0V  (r)} in Eq.\ \eqref{n = LV0}, we get 
\begin{align}
     &L(\bm r,\bm r \p,\bm r'' ,\bm r'''\!\!,z) = L^0(\bm r,\bm r \p,\bm r''\!,\bm r'''\!\!,z)  \nonumber\\
    &{+}\tfrac{1}{2}L^0(\bm r,\bm r\p,\bm r_1 ,\bm r_2,z)K(\bm r_1,\bm r_2, \bm r_3, \bm r_4)L(\bm r_3,\bm r_4,\bm r''\!,\bm r'''\!\!,z) \label{bse} .
\end{align}
Eq.\ \eqref{bse} is the Dyson equation for the interacting electron-hole propagator $L$ in the screened exchange approximation for the generalized Hxc functional (see Eq.\ (14.25) of Ref.\ \cite{martin2016interacting} or Ref.\ \cite{Rocca2010BseStern}).

% {\color{red} [Io elencherei qui tutte le proprieta' di simmetria di $L^{0}$, $K$, $L$, $V_{\scf}^{(a)}$, $n^{(a)}$ rispetto a cambio di segno dell'argomento, da $z$ a $-z$, e eventualente per operazione di complesso cognugato $^*$, che sono utili per la prossima sezione, (mettiamo solo quelle che usiamo). La prima serve a dimostrare la variazionalita', l'ultima la regola d'oro di Fermi e per calcolare i matrix elements del graphene. Per dimostrare l'ultima serve aver introdotto l'equazione di Dyson per il propagatore $L$.]
% }
We finally discuss the properties that will be used  to define and to characterize the screened formulations of the response functions.
The bare electron-hole propagator satisfies the properties:
\begin{align}
     L^{0}(\bm r, \bm r', \bm r''\!, \bm r'''\!\!,z)^* &= L^{0}(\bm r''\!, \bm r'''\!\!,  \bm r,\bm r',z^*)  \label{L0(1234 z)* = L0(3412 z*)} \\
   L^{0}(\bm r, \bm r', \bm r''\!, \bm r'''\!\!,-z) &= L^{0}(\bm r'''\!\!, \bm r''\!,  \bm r',\bm r,z) \label{L0(1234 z) = L0(4321-z)}
\end{align}
which can be proven directly from Eq.\ \eqref{L0(r)}. Instead, from the definition Eq.\ \eqref{kerneleq} follows that:
\begin{eqnarray} \label{K(1234) = K(3412)}
 K(\bm r, \bm r', \bm r''\!, \bm r''')&=&K(\bm r''\!, \bm r'''\!\!,\bm r, \bm r'),\\
 \label{K(1234) = K(4321)}
 K(\bm r, \bm r', \bm r''\!, \bm r''')&=&K(\bm r'''\!\!, \bm r''\!, \bm r', \bm r),
\end{eqnarray}
which can be proven exploiting the symmetries of the delta functions and the fact that in the SX approximation $W(\bm r,\bm r'){=}W(\bm r',\bm r)$. Combining these equations with Eq.\ \eqref{bse}, we obtain:
\begin{eqnarray}
    L(\bm r, \bm r', \bm r''\!, \bm r'''\!\!,z)^* &= L(\bm r''\!, \bm r'''\!\!,  \bm r,\bm r',z^*)  \label{L(1234 z)* = L(3412 z*)}, \\
   L(\bm r, \bm r', \bm r''\!, \bm r'''\!\!,-z) &= L(\bm r'''\!\!, \bm r''\!,  \bm r',\bm r,z). \label{L(1234 z) = L(4321-z)}
\end{eqnarray} 
{Eqs.\ \eqref{L(1234 z)* = L(3412 z*)}-\eqref{L(1234 z) = L(4321-z)} show that the symmetry properties of the bare electron hole propagator $L^{(0)}$ Eqs.\ \eqref{L0(1234 z)* = L0(3412 z*)}-\eqref{L0(1234 z) = L0(4321-z)} are inherited by the interacting propagator $L$.}

\subsection{Examples of response functions}
In the examples considered below, for simplicity we will consider a local external potential, i.e.\
\begin{equation} \label{vext is local}
    \langle\bm r|\hat{V}_\ext^{\text{nl}\,(a)}|\bm r\p\rangle{=}{V}_\ext^{(a)}(\bm r) \delta(\bm r{-}\bm r\p).
\end{equation}
At this point, we introduce the retarded response function that describes the variation of the observable tied to the perturbation $\hat V_\ext^{(a)}$ due to the presence of the perturbation $\hat V_\ext^{(b)}$. It is expressed in Fourier space as
\begin{equation} 
\label{C = v0 L v0}
C_{ab}(z) = 2V_\ext^{(a)}(\bm r_1)n^{(b)}(\bm r_1,\bm r_1, z)
\end{equation}
or equivalently, using Eq.\ \eqref{n = LV0}, as
\begin{equation} 
\label{C = v0 L v0 bis}
C_{ab}(z) = V_\ext^{(a)}(\bm r_1)L(\bm r_1,\bm r_1,\bm r_2 ,\bm r_2,z) V_{\ext}^{(b)}(\bm r_2)  .
\end{equation}
Eq.\ \eqref{C = v0 L v0 bis} is valid for a general external perturbation. Depending on the nature of the parameters $a, b$, we obtain different response functions, that correspond to different physical observables. In this Section, we present some notable examples of response functions which are of wide use in the study of molecular and crystalline systems. 

In particular, from the knowledge of the retarded interacting electron-hole propagator $L$, that is defined for clamped nuclei at the equilibrium positions, we can obtain the  phonon Green's function and the vibrational spectra (see e.g.\ \cite{maximov75,allen79,allen-book,RevModPhys.89.015003}). Notice that the clamped nuclei hypothesis corresponds to disregarding electron-phonon effects on the electron dynamics, such as the terms neglected leveraging the Migdal theorem \cite{migdal58} (cf.\  Fig.\ 1(b) of \cite{allen79}) and electron-phonon insertions in the electron Green function (cf.\ Fig.\ (1)c of \cite{allen79}).

To this purpose, for a molecule, we consider the displacements $u_{s\alpha}(t)/u_{r\beta}(t\p)$ of the atoms $s/r$ along the Cartesian coordinates $\alpha/\beta$. In Eq.\ \eqref{C = v0 L v0 bis}, we take 
\begin{align}\label{Vext = Vext finite}
V_\ext^{(a)}(\bm r_1)= \frac{\partial V_{\text{ext}}(\bm{r}_1)}{\partial u_{s\alpha}}=V^{(s\alpha)}_{\text{ext}}(\bm r_1), \\
V_\ext^{(b)}(\bm r_2)= \frac{\partial V_{\text{ext}}(\bm{r}_2)}{\partial u_{r\beta}}=V^{(r\beta)}_{\text{ext}}(\bm r_2).
\end{align}
In this case, the response function \eqref{C = v0 L v0 bis} corresponds to the nontrivial contribution to the frequency-dependent interatomic force constant matrix $C_{s\alpha,r\beta}$ \cite{giannozzi1991ab,gonze1997dynamical,calandra2010adiabatic,berges2023phonon,maximov75}, which reads
\begin{equation}
    \label{Cab phonon}
    \begin{split}
    C_{s\alpha,r\beta}(z) = V_\ext^{(s\alpha)}(\bm r_1)L(\bm r_1,\bm r_1,\bm r_2 ,\bm r_2,z) V_{\ext}^{(r\beta)}(\bm r_2).
    \end{split}
\end{equation}
The corresponding dynamical matrix $D_{s\alpha,r\beta}$ is defined as
\begin{align}\label{Dab phonon}
D_{s\alpha,r\beta}(z)=\frac{ C_{s\alpha,r\beta}(z)+ V_\ext^{(s\alpha,r\beta)}(\bm r_1) \rho\unpert(\bm r_1)}{\sqrt{M_s M_r}},
\end{align}
where $M_{s}$ represent the mass of the atom $s$.
Notice that, in order to define the dynamical matrix in Eq.~\eqref{Dab phonon} that determines the motions of the atoms, the second order term in the potential expansion in Eq.~\eqref{potentialexp} is needed, even though it enters in a term of trivial evaluation \cite{calandra2010adiabatic}. $D_{s\alpha,r\beta}$ can be used to define the displacement-displacement (divided by the nuclear masses) Green's function ~\cite{berges2023phonon} as: 
\begin{equation} \label{phonon green function}
     G^{-1}{\!\!\!\,}_{s\alpha,r\beta}(z) = z^2 - D_{s\alpha,r\beta}(z) .
\end{equation}
The phonon spectral weight is defined as:
\begin{equation}
A(\omega)={-}\lim_{\eta\rightarrow 0^+}\sum_{s\alpha}{\rm Im}\left[\frac{\omega{+}i\eta}{\pi} G_{s\alpha,s\alpha}(\omega{+}i\eta)\right] .
\end{equation}
The peaks of the spectral weight define the vibrational frequencies, while the vibrational self energy can be obtained from $D_{s\alpha,r\beta}$.

To define the density-density response in the context of periodic crystals, we consider a system with unit cell $\Omega$ and a periodic perturbation of wavevector $\bm q$. We suppose that $\bm q$ is commensurate with a Born-Von Karman supercell of dimensions $N\Omega$ which has the same periodicity of the perturbation. In this case, the density-density response function is obtained by considering Eq.\ \eqref{C = v0 L v0 bis} with 
\begin{align}\label{Vext = Vext compl}
V_\ext^{(a)}(\bm r)=e^{-i\bm q {\cdot} \bm r}, \quad V_\ext^{(b)}(\bm r)
=e^{i\bm q {\cdot} \bm r}.
\end{align}
The response in Eq.\ \eqref{C = v0 L v0 bis} now depends on $(\bm q,z)$. The density-density response function therefore reads
\begin{equation} \label{chi = cos L cos}
    \chi(\bm q,z) = \int\limits_{N\Omega} \frac{d\bm r}{N\Omega} e^{-i\bm q {\cdot} {\bm r}} L(\bm r,\bm r, \bm r_1, \bm r_1,z)e^{i\bm q {\cdot} {\bm r}_1} ,
\end{equation}
where the integral over $\bm r$ runs over the $N$ cells of the supercell and  the implicit one over $\bm r_1$ on the full crystal. In Sec.\ \ref{sec: application to graphene}, we compute the density-density response function to obtain the optical conductivity $\sigma(\bm q,z)$ of graphene. 

% Finally, notice that Eq. \eqref{Vext = Vext compl} can also represent the potential due to a monochromatic displacement of ions, in which case the response function becomes the Fourier transform of the force constants.

Finally, we also define the linear response to a monochromatic displacement of ions (i.e.\ phonons), in which the response function becomes the Fourier transform of the dynamical matrix. For the crystal case, we consider the atoms to be situated at $u^{\bm{R}}_{s\alpha}=R_{\alpha}+\tau_{s\alpha}$, $\bm{R}$ being a direct Bravais lattice vector and $\boldsymbol{\tau}_s$ the coordinate relative to the unit cell origin. We take
\begin{align}
    V_\ext^{(a)}(\bm r)&=\sum_{\bm{R}}e^{-i\bm q \cdot \left(\bm R+\boldsymbol{\tau}_s\right)}\frac{\partial V_{\text{ext}}(\bm r)}{\partial u^{\bm{R}}_{s\alpha}}{=}U_\ext^{(s\alpha,-\bm q)}(\bm r)e^{-i\bm q\cdot \bm r}, \\
    V_\ext^{(b)}(\bm r)&=\sum_{\bm{R}}e^{i\bm q \cdot \left(\bm R+\boldsymbol{\tau}_r\right)}\frac{\partial V_{\text{ext}}(\bm r)}{\partial u^{\bm{R}}_{r\beta}} 
    =U_\ext^{(r\beta,\bm q)}(\bm r)e^{i\bm q\cdot \bm r},        
\end{align}
where $U_\ext^{(r\beta,\bm q)}(\bm r_2)$ are cell-periodic potentials and the sum over $\bm R$ runs over the full crystal. Eq.\ \eqref{C = v0 L v0 bis} becomes 
\begin{align}
    C_{s\alpha,r\beta}(\bm q,z) = \frac{1}{N}\int_{N\Omega} d\bm r \,U_\ext^{(s\alpha,-\bm q)}(\bm r)e^{-i\bm q {\cdot} {\bm r}} \nonumber \\
    \times L(\bm r,\bm r, \bm r_2, \bm r_2,z)U_\ext^{(r\beta,\bm q)}(\bm r_2)e^{i\bm q {\cdot} {\bm r}_2} .
\end{align}
For any wavevector $\bm q$, the dynamical matrix and the Green's function are obtained in the same way as for Eqs.\ \eqref{Dab phonon} and \eqref{phonon green function}.

\section{Response functions in terms of screened vertices } \label{sec:sb ss s*s}
In this Section, we present three different but equivalent rewritings of the response functions in the case of a nonlocal interaction potential, starting from Eq.\ \eqref{C = v0 L v0}. The first formulation is the bare-screen, which is the most common formulation. In the static DFT framework, it is equivalent to the Sternheimer approach of Ref.~\cite{baroni2001phonons}.
Next, we introduce the screen-screen approach, extending the approach of Ref.\ \cite{calandra2010adiabatic} to the case of nonlocal interactions.
We show that this formulation is variational with respect to the electron density matrix. Thus, an approximation of the density matrix results in a quadratic error in the response function. Finally, we introduce the screen*-screen formulation, that provides an expression for the imaginary part of the response function in the form of a generalized Fermi Golden Rule, extending the approach of Ref.\ \cite{allen-book} to the case of nonlocal interactions.
We stress that the three proposed formulations are exactly equivalent. They differ only whenever one performs an approximation on the electronic density matrix, as proved numerically in Sec.\ \ref{sec: application to graphene}.

\subsection{Bare-screen response} \label{subsec: screen bare}
We start from Eq.\ \eqref{C = v0 L v0}. We use the relation between the density and the potential given by Eq.\ \eqref{n(r) = L0V  (r)}, to reformulate the response function $C_{ab}$ as
\begin{equation} \label{Cab screen bare}
    C_{ab}^{\text{bs}}(z) {=} V_{\ext}^{(a)}(\bm r_1)  L^{0}(\bm r_1, \bm r_1, \bm r_2, \bm r_3,z) V_{\scf}^{(b)}(\bm r_2,\bm r_3, z)  .
\end{equation}
Eq.\ \eqref{Cab screen bare} can be interpreted diagramatically as a response function composed by the two vertices $V_{\ext}^{(a)}$ and $V_{\scf}^{(b)}$ connected by the propagator $L^{0}$ (cf.\ Table \ref{table:diagrams}). The external potential $V_{\ext}^{(a)}$ does not depend on the electronic density matrix. The bare electron-hole propagator $L^{0}$ describes the response of noninteracting particles, as explicitly shown by Eq.\ \eqref{L0(r)}. Thus, the screening of the response due to the electronic interactions is only contained in the self-consistent potential. Accordingly, we refer to $V_{\scf}^{(b)}$ and $V_{\ext}^{(a)}$ respectively as a screened and a bare vertex. We call the formulation Eq.\ \eqref{Cab screen bare} the bare-screen response $C_{ab}^{\text{bs}}$. 

Eq.\ \eqref{Cab screen bare} is linear in the electronic density matrix. To show it, we introduce the functional 
\begin{align}
    &\mathcal{F}_{ab}[\lambda,z] = V_\ext^{(a)}(\bm r_1) L^{0}(\bm r_1,\bm r_1,\bm r_2,\bm r_3,z) \nonumber\\
     &\times \left[V_{\ext}^{(b)}(\bm r_2)\delta(\bm r_2{-}\bm r_3) {+} K(\bm r_2, \bm r_3, \bm r_4, \bm r_5)\lambda(\bm r_4, \bm r_5)\right],        \label{F sb}
\end{align}    
that satisfies the relation 
\begin{equation} \label{Csb = Fab}
    C_{ab}^{\text{bs}}(z) = \mathcal{F}_{ab}[\lambda {=}  n^{(b)}(z),z] .
\end{equation}
From Eq.\ \eqref{F sb}, it follows that
\begin{align}
&\frac{\delta \mathcal{F}_{ab}[\lambda,z]}{\delta \lambda(\bm r_i,\bm r_j)}\eval_{\lambda {=} n^{(b)}(z)}\!\!\!=\nonumber\\
&V_\ext^{(a)}(\bm r_1) L^{0}(\bm r_1,\bm r_1,\bm r_2,\bm r_3,z)K(\bm r_2, \bm r_3, \bm r_i, \bm r_j). \label{dbsdn}
\end{align}
The above derivative does not depend on the electronic density matrix. Therefore, an error of $\delta n^{(b)}(\bm r, \bm r\p,z)$ on the density matrix results in an error of order $\delta n^{(b)}(\bm r, \bm r\p,z)$ in the bare-screen response, thus proving its linearity with respect to changes in the electronic density matrix.

\subsection{Screen-screen response } \label{sec: variational}
\label{subsec: screen screen}
To obtain the screen-screen response we manipulate Eq.\ \eqref{Cab screen bare} by expressing the external potential $V_{\ext}^{(a)}$ in terms of the self-consistent potential $V_{\scf}^{(a)}$. 
Therefore, we rewrite Eq.\ \eqref{Cab screen bare} using the relation between the screened potential and the external potential, Eq.\ \eqref{V_scf(r) = V + Kn}. $C_{ab}(z)$ then becomes
\begin{align}
     &C_{ab}(z){=}\bigl[V_{\scf}^{(a)}(\bm r_2,\bm r_1, {-}z){-}K(\bm r_2, \bm r_1, \bm r_3, \bm r_4) \nonumber \\ &\times n^{(a)}(\bm r_3, \bm r_4,{-}z)\bigr]
   L^{0}(\bm r_1, \bm r_2, \bm r_5, \bm r_6,z) V_{\scf}^{(b)}(\bm r_5,\bm r_6, z). \label{Cab = [v -kn]n}
\end{align}
Further, we use the connection between the bare electron-hole propagator $L^{0}$ and the induced density, given by Eq.\ \eqref{n(r) = L0V  (r)}, and the symmetry properties of the kernel expressed in Eqs.\ \eqref{K(1234) =  K(3412)} and \eqref{K(1234) = K(4321)}. With these ingredients, Eq.\ \eqref{Cab = [v -kn]n} is rewritten as 
\begin{align}
    &C_{ab}^{\text{ss}}(z){=}  V_{\scf}^{(a)}(\bm r_2,\bm r_1,{-}z) L^{0}(\bm r_1,\bm r_2,\bm r_3,\bm r_4,z)V_{\scf}^{(b)}(\bm r_3,\bm r_4,z) \nonumber\\
    &{-}2n^{(a)}(\bm r_2,\bm r_1,{-}z)K(\bm r_1,\bm r_2,\bm r_3,\bm r_4)n^{(b)}(\bm r_3,\bm r_4,z).\label{Cab screen screen}
\end{align}
The first term in Eq.\ \eqref{Cab screen screen} is made up of two screened vertices $ V_{\scf}^{(a)},  V_{\scf}^{(b)}$, connected by the bare electron-hole propagator $L^{0}$. The second term in Eq.\ \eqref{Cab screen screen} removes double-counting terms that occur in the first term (cfr.\ Table \ref{table:diagrams} for the diagramatic representation). We refer to the formulation in Eq.~\eqref{Cab screen screen} as the screen-screen response function $C^{\text{ss}}_{ab}(z)$.

Eq.\ \eqref{Cab screen screen} represents the generalization of the result obtained in Ref.\ \cite{calandra2010adiabatic} in the case of a nonlocal interaction potential, and is the main result of this work. Closely to the result obtained in the aforementioned work, we show in the following section that the response function in Eq.~\eqref{Cab screen screen} is variational with respect to the density matrix. 

\begingroup
\setlength{\tabcolsep}{12pt}
\renewcommand{\arraystretch}{2}
\begin{table*}[t] 
    \centering
    \begin{tabular}{|c c c|}
    \hline
    \raisebox{0.2\height}{Eq.} & \raisebox{0.2\height}{Expression} & \raisebox{0.2\height}{Diagramatic representation}  \\
    \hline
    \multicolumn{3}{|c|}{\rule{0pt}{0.0cm}}\\
    \eqref{V_scf(r) = V + Kn} & $V^{(b)}_{\scf} = V^{(b)}_{\ext}+ Kn^{(b)}$&\includegraphics[keepaspectratio,valign=c,height = 0.9cm]{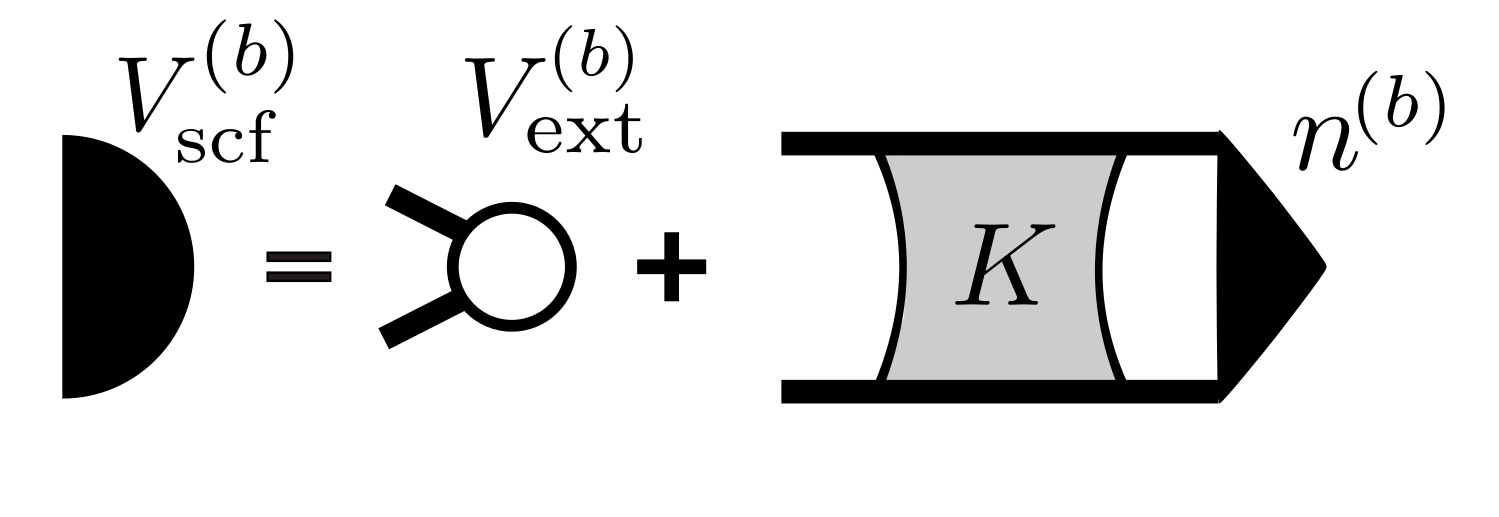} \\[1cm]
     \eqref{kerneleq} & $K = 2f_{\Hxc}-W$&\includegraphics[keepaspectratio,valign=c,height = 0.6cm]{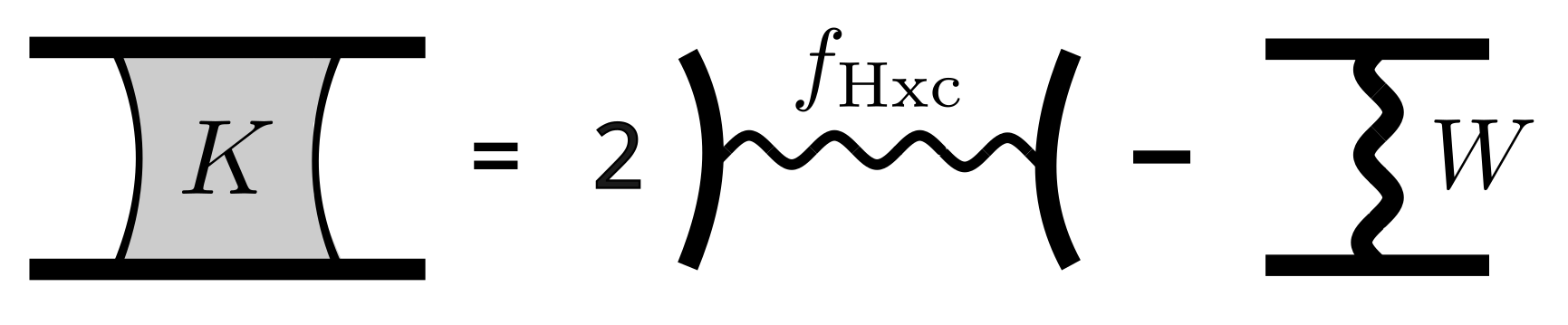} \\[1cm]
    \eqref{n(r) = L0V  (r)} & $2n^{(b)} = L^{0}V_{\scf}^{(b)}$&\includegraphics[keepaspectratio,valign=c,height = 0.8cm]{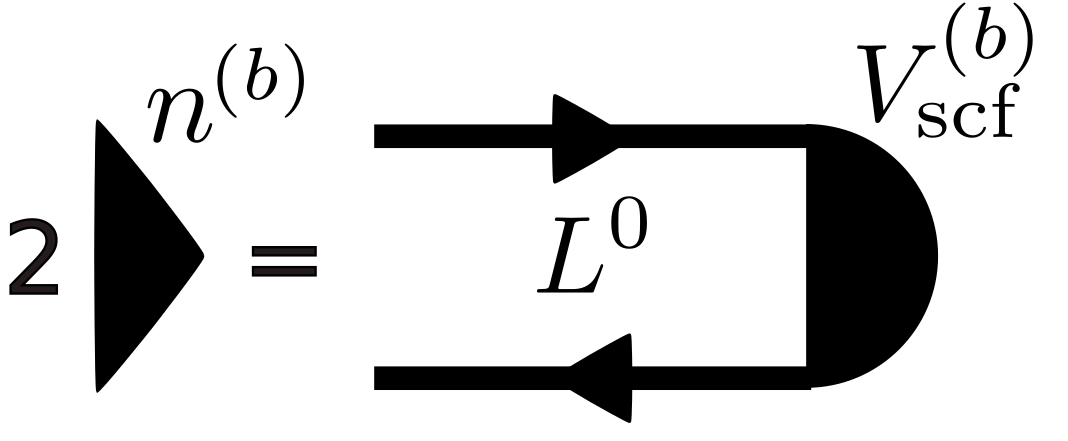} \\[1cm]
         \eqref{Cab screen bare} & $C^{\text{bs}}_{ab} = V_{\ext}^{(a)}L^{0}V_{\scf}^{(b)}$ & \includegraphics[keepaspectratio,valign=c,height = 0.8cm]{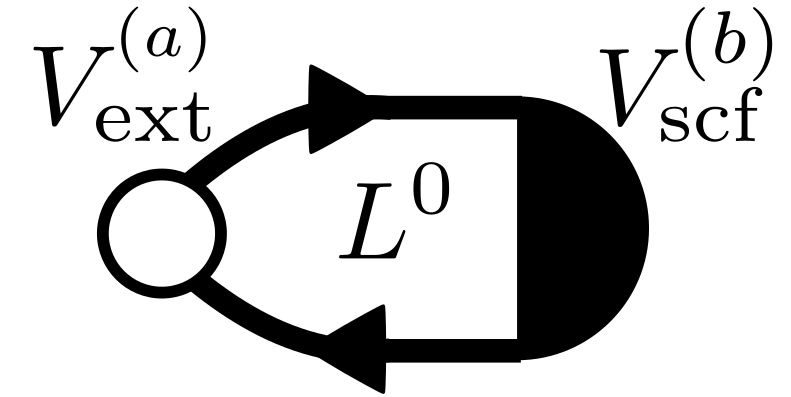} \\[1cm]
         \eqref{Cab screen screen} & $C^{\text{ss}}_{ab} = V^{(a)}_{\scf}L^{0}V^{(b)}_{\scf} - 2n^{(a)} K n^{(b)}$ &  $\includegraphics[keepaspectratio,valign=c,height = 0.8cm]{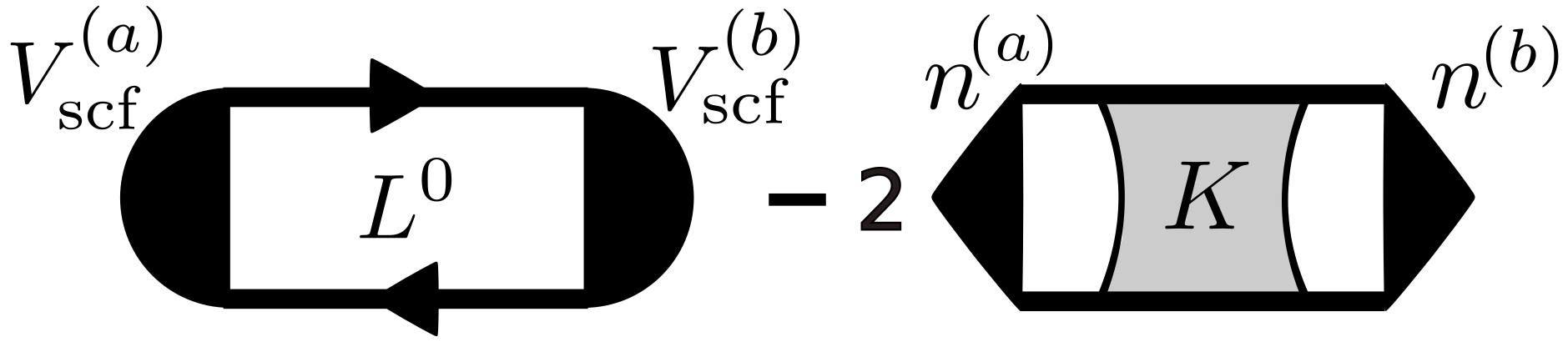}  $  \\[1cm]
         \eqref{Cab screen* screen z*} & $  C^{\text{s*s}}_{ab}= V^{(a)*}_{\scf}L^{0}V^{(b)}_{\scf} - 2n^{(a)*}Kn^{(b)}$ & $\includegraphics[keepaspectratio,valign=c,height = 0.8cm]{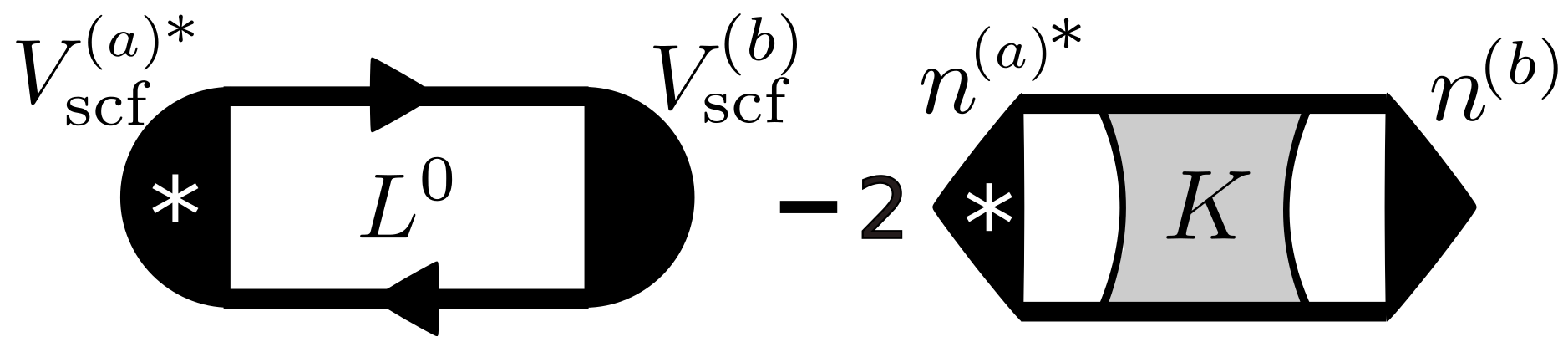} $    \\[1cm]
    \hline
    \end{tabular}
    \caption{Diagramatic representation of the the self-consistent equations for the screened vertex, kernel, density matrix and different formulations of the response functions. The full black semicircle represent the screened vertex $V_{\scf}$, while an empty circle symbolizes the bare (local) vertex $V_\ext$. The bare electron-hole propagator $L^{0}$ is represented as two lines, while the shaded grey block is the interaction kernel $K$. The density matrix is sketched with a full black triangle. The star symbolizes complex conjugation. The Hartree plus exchange-correlation kernel $f_{\Hxc}$ and screened Coulomb interaction $W$ are represented by thin and thick wavy lines respectively. Notice that the diagrams contained in the interaction kernel depend on the particular approximation scheme adopted. In the case of the SX approximation, $f_{\Hxc} = v$, where $v$ is the bare Coulomb interaction, and $W = \epsilon^{-1}v$ is the Coulomb interaction screened by the RPA inverse dielectric constant.}
    \label{table:diagrams}
\end{table*}
\endgroup

\subsubsection{Stationary response functional}\label{stationary proof}
To demonstrate the variationality of the screen-screen response, we introduce the response functional 
\begin{widetext}

%%% ==== Gio === %%%%
% ho pensato di mettere tutto in widetext questo paragrafo perchè spezzare tre volte con equazioni lunghe fa un po di confusione con l'allineamento del testo (alcune mezze colonne sono vuote)
%%%%%%%%%%%%%%%%%%%%%%%
\begin{align}
     &F_{ab}[\lambda,\lambda\p,z] {=}\! \left[ V_{\ext}^{(a)}(\bm r_1)\delta(\bm r_1{-}\bm r_2) {+} K(\bm r_2, \bm r_1, \bm r_5, \bm r_6)\lambda(\bm r_5, \bm r_6) \right] L^{0}(\bm r_1,\bm r_2,\bm r_3,\bm r_4,z) \nonumber\\
     &\times \left[V_{\ext}^{(b)}(\bm r_3)\delta(\bm r_3{-}\bm r_4) {+} K(\bm r_3, \bm r_4, \bm r_7, \bm r_8)\lambda\p(\bm r_7, \bm r_8)\right] - 2\lambda(\bm r_2,\bm r_1) K(\bm r_1,\bm r_2,\bm r_3,\bm r_4) \lambda\p(\bm r_3,\bm r_4)  ,\label{F[rho,rho']}
\end{align}  
that satisfies the relation
\begin{equation}
    \label{Css = Fab}
    C_{ab}^{\text{ss}}(z) = F_{ab}[\lambda {=} n^{(a)}({-}z),\lambda\p {=} n^{(b)}(z),z].
\end{equation}
The functional $F_{ab}[\lambda,\lambda\p,z]$ in Eq.\  \eqref{F[rho,rho']} satisfies
\begin{equation}
\begin{split} \label{dF = 0}
    \frac{\delta F_{ab}[\lambda,\lambda\p,z]}{\delta \lambda(\bm r_i, \bm r_j)}\eval_{\substack{\lambda =n^{(a)}({-}z),\\ \lambda\p =n^{(b)}(z)} } = 0 ,\quad
    \frac{\delta F_{ab}[\lambda,\lambda\p,z]}{\delta \lambda\p(\bm r_i, \bm r_j)}\eval_{\substack{\lambda =n^{(a)}({-}z),\\ \lambda\p =n^{(b)}(z)} } = 0.
\end{split}
\end{equation}
In fact, let us consider the derivative of $F_{ab}[\lambda,\lambda\p,z]$ with respect to $\lambda$, which reads
\begin{equation} \label{dF/drho pre}
     \frac{\delta F_{ab}[\lambda,\lambda\p,z]}{ \delta \lambda(\bm r_i, \bm r_j)}\eval_{\substack{\lambda {=} n^{(a)}({-}z)\\ \lambda\p{=}n^{(b)}(z)}} \!\!\!=K(\bm r_2,\bm r_1,\bm r_i,\bm r_j)L^{0}(\bm r_1,\bm r_2,\bm r_3,\bm r_4,z)V_{\scf}^{(b)}(\bm r_3,\bm r_4,z)- K(\bm r_j,\bm r_i,\bm r_1,\bm r_2)2n^{(b)}(\bm r_1,\bm r_2,z).
\end{equation}
Using Eq.\ \eqref{n = LV0}, we  recast Eq.\ \eqref{dF/drho pre} as 
\begin{equation} \label{dF/drho}
        \frac{\delta F_{ab}[\lambda,\lambda\p,z]}{ \delta \lambda(\bm r_i, \bm r_j)}\eval_{\substack{\lambda {=} n^{(a)}({-}z)\\ \lambda\p{=}n^{(b)}(z)}} \!\!\!=\left[K(\bm r_2,\bm r_1,\bm r_i,\bm r_j)- K(\bm r_j,\bm r_i,\bm r_1,\bm r_2)\right]2n^{(b)}(\bm r_1,\bm r_2,z) =0 .
\end{equation}
where the term in square brackets in Eq.\ \eqref{dF/drho} is zero due to the symmetries of the kernel expressed in Eqs.\ \eqref{K(1234) = K(3412)} and \eqref{K(1234) = K(4321)}. Instead, the derivative with respect $\lambda\p$ reads
\begin{equation} \label{dF/drho 2}
        \frac{\delta F_{ab}[\lambda,\lambda\p,z]}{ \delta \lambda\p(\bm r_i, \bm r_j)}\eval_{\substack{\lambda {=} n^{(a)}({-}z)\\ \lambda\p{=}n^{(b)}(z)}} \!\!\!=\left[V_{\scf}^{(a)}(\bm r_2,\bm r_1,{-}z)L^{0}(\bm r_1,\bm r_2,\bm r_3,\bm r_4,z)- 2n^{(a)}(\bm r_4,\bm r_3,{-}z)\right] K(\bm r_3,\bm r_4,\bm r_i,\bm r_j)= 0 .
\end{equation}
\end{widetext}
The term in square brackets of Eq.\ \eqref{dF/drho 2} vanishes using the symmetry property \eqref{L0(1234 z) = L0(4321-z)} of the bare electron-hole propagator $L^0$ in its relation with the density matrix \eqref{n(r) = L0V  (r)}.
% ==== TR assumption : OLD ========%
% The term in square brackets of Eq. \eqref{dF/drho 2} is zero assuming that the Hamiltonian, in absence of the external potential, $\hat H\unpert$, satisfies time-reversal symmetry. In such case, the unperturbed eigenfunctions  can be chosen to be real, i.e.\ $\psi\unpert_{i}(\bm r){=}\psi\unpert_{i}(\bm r)^*$. Then, the property 
% \begin{align}\label{L01m}
% L^{0}(\bm r, \bm r', \bm r''\!, \bm r'''\!\!,z) {=}  L^{0}(\bm r''\!, \bm r'''\!\!, \bm r, \bm r',z)
% \end{align}
% holds, and the term in square brackets vanishes due to Eq. \eqref{n = LV0}.

Eq.\ \eqref{dF = 0}  implies that the screen-screen response function of Eq.\ \eqref{Cab screen screen} is an extremal point of the functional $F_{ab}[\lambda,\lambda\p,z]$, in both $\lambda$ and $\lambda\p$. In other words, an error of $\delta n^{(b/a)}(\bm r, \bm r\p,\pm z)$ on the density matrix results in an error of order $\left[\delta n^{(b/a)}(\bm r, \bm r\p,\pm z)\right]^2$ in the screen-screen response.

\subsection{Screen*-screen response} \label{subsec: fgr}
In Sec.\ \ref{subsec: screen bare} and \ref{subsec: screen screen} we have discussed respectively the definitions of the bare-screen and screen-screen response functions. As anticipated in the introduction, we now introduce another possible formulation of the electronic response, from which a generalization of the Fermi Golden Rule for the imaginary part of the response function can be demonstrated.

Performing the same passages as for the screen-screen response, but with ${-}z^*$ instead of ${-}z$, we rewrite the response function of Eq.\ \eqref{C = v0 L v0} as
\begin{align}
  & C^{\text{s*s}}_{ab}(z) {=}  V_{\scf}^{(a)}(\bm r_2,\bm r_1,\!{-}z^*) L^{0}(\bm r_1,\bm r_2,\bm r_3,\bm r_4,z)V_{\scf}^{(b)}(\bm r_3,\bm r_4,z) \nonumber \\
    &{-}2n^{(a)}(\bm r_2,\bm r_1,{-}z^*)K(\bm r_1,\bm r_2,\bm r_3,\bm r_4)n^{(b)}(\bm r_3,\bm r_4,z) \label{Cab screen* screen z*}  .
\end{align}

% now indicated for brevity with $F_{\bm q}=F_{-\bm q\bm q}$
It is interesting to notice that Eq.\ \eqref{Cab screen* screen z*} determines the holomorphic quantity $C^{\text{s*s}}_{ab}(z)$ by summing two separately nonholomorphic terms.   The screen*-screen response in Eq.~\eqref{Cab screen* screen z*} can be obtained from the functional in Eq.~\eqref{F[rho,rho']}, as
\begin{equation}
    \label{Cs*s = Fab}
    C^{\text{s*s}}_{ab}(z) = F_{ab}[\lambda {=} n^{(a)}({-}z^*),\lambda\p {=} n^{(b)}(z),z].
\end{equation}
However, $C^{\text{s*s}}_{ab}(z)$ is not a stationary point of the functional $F[\lambda,\lambda\p,z]$: 
\begin{widetext}
\begin{equation} \label{dF/drho' s*s}
    \begin{split}
        \frac{\delta F_{ab}[\lambda,\lambda\p,z]}{\delta \lambda\p(\bm r_i,\bm r_j)}\eval_{\substack{\lambda {=} n^{(a)}({-}z^*)\\ \lambda\p{=}n^{(b)}(z)}} \!\!\!{=}K(\bm r_3, \bm r_4, \bm r_i,\bm r_j)\left[V_{\scf}^{(a)}(\bm r_2, \bm r_1,{-}z^*)L^{0}(\bm r_1, \bm r_2, \bm r_3, \bm r_4,z)- 2n^{(a)}(\bm r_4, \bm r_3,{-}z^*)\right].
    \end{split}
\end{equation}
\end{widetext}
The term in square brackets in Eq.\ \eqref{dF/drho' s*s}  is zero only in the nonresonant case in which $\eta{=}0, z {=}\omega$ and the screen-screen and screen*-screen formulations coincide. Therefore, an error of $\delta n^{(b)}(\bm r, \bm r\p,z)$ on the density matrix results in an error of order $\delta n^{(b)}(\bm r, \bm r\p,z)$ in the screen*-screen response, thus proving its linearity with respect to changes in the electronic density matrix.

\subsubsection{Generalized Fermi Golden Rule}
Let us consider the particular case where:
\begin{align}
V_{\ext}^{(a)}(\bm r) = V_{\ext}^{(b)}(\bm r)^*.
\end{align}
This correspond to the diagonal response function of  a perturbation, either real, or modulated by a wavevector $\bm q$, as that of a phonon with quasimomentum $\hbar \bm q$.
For the present case, the properties Eqs.\ \eqref{L(1234 z)* = L(3412 z*)}-\eqref{L(1234 z) = L(4321-z)} and Eqs.\ \eqref{L0(1234 z)* = L0(3412 z*)}-\eqref{L0(1234 z) = L0(4321-z)}  respectively of the interacting and bare electron-hole propagators $L$ and $L^0$ can be leveraged in the definitions of the density matrix Eq.\ \eqref{n = LV0} and Eq.\ \eqref{n(r) = L0V  (r)} to show that
\begin{align}
n^{(a)}(\bm r',\bm r,{-}z^*)&=n^{(b)}(\bm r,\bm r',z)^*, \label{n^a* = nb}\\
V_{\scf}^{(a)}(\bm r', \bm r,{-}z^*) &= V_{\scf}^{(b)}(\bm r, \bm r',z)^*. \label{Va^* = Vb}
\end{align}
Using Eq.\ \eqref{n^a* = nb} and Eq.\ \eqref{Va^* = Vb}, and the properties Eq.\ \eqref{K(1234) = K(3412)} of $K$ and Eq.\ \eqref{L(1234 z)* = L(3412 z*)} of $L^0$, we obtain:
\begin{align}\label{Im  C}
 &\Im  C^{\text{s*s}}_{ab}(z) {=}V_{\scf}^{(b)}(\bm r_1,\bm r_2,z)^* V_{\scf}^{(b)}(\bm r_3,\bm r_4,z) \nonumber\\
& \times \frac{1}{2i} [L^{0}(\bm r_1,\bm r_2,\bm r_3,\bm r_4,z) {-} L^{0}(\bm r_1,\bm r_2,\bm r_3,\bm r_4,z^*)],
\end{align}
Passing from the real space representation to the Dirac one, we get
\begin{align}
\Im C^{\text{s*s}}_{ab}(z) {=} 2 \sum_{ij} |\langle \psi^{(0)}_i|\hat V^{(b)}_{\text{scf}}(z)|\psi^{(0)}_j \rangle|^2   \nonumber\\ 
\times \Im \left[
\frac{f^{(0)}_j-f^{(0)}_i}{\hbar z {-} (\varepsilon_i\unpert {-} \varepsilon_j\unpert)}
\right]. \label{ImCs*s}
\end{align}
Supposing that the matrix elements $\langle \psi^{(0)}_i|\hat V^{(b)}_{\text{scf}}(z)|\psi^{(0)}_j \rangle$ remain finite in the $\eta {\to}0^+$ limit, and that $C^{\text{s*s}}_{ab}(z)$ represents a self-energy, it  follows that
\begin{align}
        \Gamma(\omega) {=} {-}\frac{1}{\hbar}\lim\limits_{\eta {\to}0^+}2\Im C^{\text{s*s}}_{ab}(z) {=}\frac{2\pi}{\hbar}2\sum_{ij}(f^{(0)}_j{-} f^{(0)}_i) \nonumber \\
        \times \bigl|\langle\psi\unpert_i|\hat V_{\scf}^{(b)}(\omega)|\psi_{j}\unpert\rangle\bigr|^2\delta(\varepsilon\unpert_i {-} \varepsilon\unpert_j{-}\hbar \omega  )  . \label{fermi golden rule}
\end{align}
Eq.\ \eqref{fermi golden rule} consists in the generalization of the Fermi Golden Rule \cite{fermi1950nuclear} to the case of self-consistent Hamiltonians. When computed e.g.\ at the frequency of the phonon, it corresponds to the full width at half maximum and it is widely used to compute phonon scattering rates.
The combination of the square modulus of the screened vertices, the Fermi-Dirac populations, and the Dirac delta in Eq.\ \eqref{fermi golden rule} clearly makes $\Gamma(\omega)$ a positive definite quantity. This property allows for a probabilistic interpretation of $\Gamma$; much like in semiclassical Boltzmann-Equation approaches, the imaginary part of the response function associated with dissipation can be interpreted in terms of scattering processes. 

As for the standard Fermi Golden Rule, Eq.\ \eqref{fermi golden rule} is only valid when the summation over states $i,j$ forms a continuum, and $\omega$ is taken within this continuum. In solids this is generally obtained considering a Born–von Karman periodic cell of volume $V{=}N\Omega$ and performing the $V{\to}\infty$ thermodynamic limit before the $\eta {\to}0^+$ one. 
Moreover, assuming that the matrix elements of the screened potential are finite, to obtain  Eq.\ \eqref{fermi golden rule} we neglected the contributions from the electronic excitations generated by the electron-electron interaction decoupled from an underlined continuum, such as undamped plasmons or bounded excitons. In this cases, $\langle\psi\unpert_i|\hat V_{\scf}^{(b)}(\omega)|\psi_{j}\unpert\rangle$ diverges for $\eta{\to}0^+$ at resonance, even in the thermodynamic limit. These contributions are instead fully included in Eq.\ \eqref{ImCs*s}, that remains valid in the $\eta{\to}0^+$ also in presence of undamped plasmons or excitons. 

It is important to stress that the bare-screen [see Eq.\ \eqref{Cab screen bare}], the screen-screen [see Eq.\ \eqref{Cab screen screen}] and the screen*-screen [see Eq.\ \eqref{Cab screen* screen z*}] are equivalent formulations of the same physical observable.
Thus, the generalized Fermi Golden Rule holds for each formulation as $\Im C^{\text{s*s}}_{ab}(z) {=} \Im C^{\text{ss}}_{ab}(z) {=} \Im C^{\text{bs}}_{ab}(z)$, where
the $\eta {\to} 0^+$ limit is given by Eq.\ \eqref{fermi golden rule}, when the screened matrix elements remain finite. 

\section{Response functions in terms of partially screened vertices} \label{sec: pp}

So far, we expressed the linear response in term of either  the external unscreened potential, $V_\text{ext}^{(b)}$, or the interacting, fully screened potential, $V_\text{scf}^{(b)}$. However, there are instances where it is advantageous to introduce and use a partially screened potential, that we call $V_\text{p}^{(b)}$, where the $V_\text{ext}^{(b)}$ is dressed by only a part of the interacting  kernel $K$. In this section,  we reformulate the response function in terms of partially screened vertices and propagators. 
% we neglected the possible presence of a retardation effect in the interaction kernel $K$. Moreover,
{\color{gray}

}
\subsection{Partial screening of the vertices } \label{subsec: partial definitions}

We introduce partial screening effects, by partitioning  the kernel of the interaction in Eq.\ \eqref{kernel retarded} as: 
\begin{equation}
    \label{k(z) = k1 + k2}
    K(\bm r, \bm r\p,\bm r\pp, \bm r{'''}){=}K^{\text{p}}(\bm r, \bm r\p,\bm r\pp, \bm r{'''}) {+} K^{\text{c}}(\bm r, \bm r\p,\bm r\pp, \bm r{'''})
\end{equation}
assuming that the partial kernel  $K^\text{p}$ is real (as $K$ and thus $K^{\text{c}}$) and has the same symmetry properties
\begin{eqnarray} \label{Kp(1234) = Kp(3412)}
 K^{\text{p}}(\bm r, \bm r', \bm r''\!, \bm r''')&=&K^{\text{p}}(\bm r''\!, \bm r'''\!\!,\bm r, \bm r'),\\
 \label{Kp(1234) = Kp(4321)}
 K^{\text{p}}(\bm r, \bm r', \bm r''\!, \bm r''')&=&K^{\text{p}}(\bm r'''\!\!, \bm r''\!, \bm r', \bm r)  .
\end{eqnarray}
From these properties of $K^{\text{p}}$ and the analogous ones of $K$, it follows that:
\begin{eqnarray} \label{Kc(1234) = Kc(3412)}
 K^{\text{c}}(\bm r, \bm r', \bm r''\!, \bm r''')&=&K^{\text{c}}(\bm r''\!, \bm r'''\!\!,\bm r, \bm r'),\\
 \label{Kc(1234) = Kc(4321)}
 K^{\text{c}}(\bm r, \bm r', \bm r''\!, \bm r''')&=&K^{\text{c}}(\bm r'''\!\!, \bm r''\!, \bm r', \bm r) .
\end{eqnarray}
Using the repartion of the interaction in Eq.\ \eqref{k(z) = k1 + k2}, we define a partially screened potential 
\begin{align}
V_\text{p}^{(b)}(\bm r,\bm r\p, z) &= V^{\text{nl}\,(b)}_{\ext}(\bm r,\bm r\p)\nonumber \\
&{+}K^{{\text{p}}}(\bm r, \bm r\p, \bm r_1, \bm r_2) n^{(b)}(\bm r_1, \bm r_2,z). \label{Vp(r) = V + Kpn}
\end{align}
Using Eq.\ \eqref{k(z) = k1 + k2}, we get from Eq.\ \eqref{V_scf(r) = V + Kn} the relation between the self-consistent induced potential and the partially screened potential
\begin{align}
    V_{\text{scf}}^{(b)}(\bm r,\bm r\p,z) &= V_{\text{p}}^{(b)}(\bm r,\bm r\p,z) \nonumber \\&+K^{\text{c}}(\bm r, \bm r\p, \bm r_1, \bm r_2) n^{(b)}(\bm r_1, \bm r_2,z). \label{Vscf = Vp + Kcn}
\end{align}
In analogy with Eqs.\ \eqref{n = LV0} and \eqref{n(r) = L0V  (r)}, we define the induced density matrix in terms of a  partially interacting propagator $L^{\text{c}}$ 
\begin{equation} \label{n = Lc vp}
        2n^{(b)}(\bm r,\bm r\p,z) {=}  L^{\text{c}}(\bm r, \bm r\p, \bm r_1, \bm r_2,z) V_{\text{p}}^{(b)}(\bm r_1,\bm r_2, z) . 
\end{equation}
To elucidate its differences with the interacting electron-hole propagator $L$ defined in Eq.\ \eqref{bse}, we derive the Dyson equation of the partially interacting propagator $L^{\text{c}}$. Using Eq.\ \eqref{Vscf = Vp + Kcn} in the definition of the density matrix in terms of the self-consistent potential Eq.\ \eqref{n(r) = L0V  (r)}, we get 
\begin{align}
    &2n^{(b)}(\bm r,\bm r\p,z) = L^{0}(\bm r, \bm r\p, \bm r_1, \bm r_2,z)\nonumber \\& \times\left[ V_{\text{p}}^{(b)}(\bm r_1,\bm r_2,z) 
+K^{\text{c}}(\bm r_1, \bm r_2, \bm r_3, \bm r_4) n^{(b)}(\bm r_3, \bm r_4,z) \right] . \label{nb =L0 Vp + Kc n}
\end{align}
By rewriting the density in terms of partially screened quantities as in Eq.\ \eqref{n = Lc vp}, we recast Eq.\ \eqref{nb =L0 Vp + Kc n} as
\begin{align}
    &L^{\text{c}}(\bm r, \bm r\p, \bm r_1, \bm r_2,z) V_{\text{p}}^{(b)}(\bm r_1,\bm r_2, z) =\nonumber\\
    &L^{0}(\bm r, \bm r\p, \bm r_1, \bm r_2,z)\bigl[ \delta (\bm r_1{-}\bm r_5)\delta (\bm r_2 {-} \bm r_6) \nonumber \\
&+\tfrac{1}{2}K^{\text{c}}(\bm r_1, \bm r_2, \bm r_3, \bm r_4) L^{\text{c}}(\bm r_3, \bm r_4, \bm r_5, \bm r_6,z)  \bigr] V_{\text{p}}^{(b)}(\bm r_5,\bm r_6,z)
 ,
\end{align}
from which it follows that
\begin{align}
    &L^{\text{c}}(\bm r, \bm r', \bm r''\!, \bm r'''\!\!,z) = L^{0}(\bm r, \bm r', \bm r''\!, \bm r'''\!\!,z) \nonumber \\
    &{+}\tfrac{1}{2}L^0(\bm r,\bm r\p,\bm r_1 ,\bm r_2,z)K^{\text{c}}(\bm r_1,\bm r_2, \bm r_3, \bm r_4)L^{\text{c}}(\bm r_3,\bm r_4,\bm r'' ,\bm r'''\!\!,z) . \label{Lc dyson}
\end{align}
Eq.\ \eqref{Lc dyson} is the Dyson equation for the partially interacting propagator. 

For completeness, we also express the relation between the partially interacting propagator $L^{\text{c}}$ and the fully interacting propagator $L$. We plug the relation Eq.\ \eqref{Vp(r) = V + Kpn} in Eq.\ \eqref{n = Lc vp} , obtaining
\begin{align}
    &2n^{(b)}(\bm r,\bm r\p,z) = L^{\text{c}}(\bm r, \bm r\p, \bm r_1, \bm r_2,z)\nonumber \\& \times\left[ V_{\text{ext}}^{(b)}(\bm r_1,\bm r_2,z) 
+K^{\text{p}}(\bm r_1, \bm r_2, \bm r_3, \bm r_4) n^{(b)}(\bm r_3, \bm r_4,z) \right] . \label{nb =Lc V0 + Kp n}
\end{align}
Expressing the density matrix in Eq.\ \eqref{nb =Lc V0 + Kp n} in terms of the interacting electron-hole propagator $L$ from Eq.\ \eqref{n = LV0}, we get
\begin{align}
    &L(\bm r, \bm r\p, \bm r_1, \bm r_2,z) V_{\text{ext}}^{(b)}(\bm r_1,\bm r_2, z) =\nonumber\\
    &L^{\text{c}}(\bm r, \bm r\p, \bm r_1, \bm r_2,z)\bigl[ \delta (\bm r_1{-}\bm r_5)\delta (\bm r_2 {-} \bm r_6) \nonumber \\
&+\tfrac{1}{2}K^{\text{p}}(\bm r_1, \bm r_2, \bm r_3, \bm r_4) L(\bm r_3, \bm r_4, \bm r_5, \bm r_6,z)  \bigr] V_{\text{ext}}^{(b)}(\bm r_5,\bm r_6,z)
 ,
\end{align}
from which it follows that 
\begin{align}
    &L(\bm r, \bm r', \bm r''\!, \bm r'''\!\!,z) = L^{\text{c}}(\bm r, \bm r', \bm r''\!, \bm r'''\!\!,z) \nonumber \\
    &{+}\tfrac{1}{2}L^{\text{c}}(\bm r,\bm r\p,\bm r_1 ,\bm r_2,z)K^{\text{p}}(\bm r_1,\bm r_2, \bm r_3, \bm r_4)L(\bm r_3,\bm r_4,\bm r'' ,\bm r'''\!\!,z) . \label{L = Lc + Kp} 
\end{align}
From Eq.\ \eqref{L = Lc + Kp} it is clear that to obtain the fully interacting electron-hole propagator $L$, one must incorporate into the partially interacting propagator $L^{\text{c}}$ the remainder of the interaction comprised in $K^{\text{p}}$.

Using the properties of the bare propagator expressed in Eqs.\ \eqref{L0(1234 z)* = L0(3412 z*)}-\eqref{L0(1234 z) = L0(4321-z)} and of the complementary part of the retarded kernel $K^\text{c}$ in Eqs.\ \eqref{Kc(1234) = Kc(3412)}-\eqref{Kc(1234) = Kc(4321)}, it follows from Eq.\ \eqref{Lc dyson} that the partially interacting propagator satisfies
\begin{align}
   L^{\text{c}}(\bm r, \bm r', \bm r''\!, \bm r'''\!\!,z)^* &= L^{\text{c}}(\bm r''\!, \bm r'''\!\!,  \bm r,\bm r',z^*)  \label{Lc(1234 z)* = Lc(3412 z*)} \\
   L^{\text{c}}(\bm r, \bm r', \bm r''\!, \bm r'''\!\!,-z) &= L^{\text{c}}(\bm r'''\!\!, \bm r''\!,  \bm r',\bm r,z)  \label{Lc(1234 z) = Lc(4321-z)}
\end{align}

%The definition of the partially screened potential in Eq.\ \eqref{Vp(r) = V + Kpn} is analogous to the definition of the self-consistent potential in Eq.\ \eqref{V_scf(r) = V + Kn}. 

%The definition of the partially screened potential in Eq.\ \eqref{Vp(r) = V + Kpn} features the full induced denisty matrix as $n^{(b)}$, as per the self-consistent potential defined in Eq.\ \eqref{V_scf(r) = V + Kn}. However, 

\subsection{Partial screen - partial screen response}
As done in Sec.\ \ref{sec:sb ss s*s}, we  express the response functions in terms of the partially screened quantities defined in Sec.\ \ref{subsec: partial definitions}.  First, we assume for simplicity that the external potential in Eq.\ \eqref{Vp(r) = V + Kpn} is local, i.e.\ satisfies Eq.\ \eqref{vext is local}. In this case, the response function $C_{ab}(z)$ is expressed as in Eq.\ \eqref{C = v0 L v0}. Using \eqref{n = Lc vp}, we recast the response function as 
\begin{equation} 
\label{C = v0 Lp vc}
C_{ab}(z) = V_\ext^{(a)}(\bm r_1)L^{\text{c}}(\bm r_1, \bm r_1, \bm r_2, \bm r_3,z) V_{\text{p}}^{(b)}(\bm r_2,\bm r_3, z)  .
\end{equation}

To obtain the partially screened-partially screened response, we manipulate Eq.\ \eqref{C = v0 Lp vc} by expressing the external potential $V_{\ext}^{(a)}$ in terms of the partially screened potential $V_{\text{p}}^{(a)}$. 
Therefore, we rewrite Eq.\ \eqref{C = v0 Lp vc} using the relation between the partially screened potential and the external potential, Eq.\ \eqref{Vp(r) = V + Kpn}. $C_{ab}(z)$ then becomes
\begin{align}
   C_{ab}(z)&{=}\bigl[V_{\text{p}}^{(a)}(\bm r_2,\bm r_1, {-}z) \nonumber \\ 
   & {-}K^{\text{p}}(\bm r_1, \bm r_2, \bm r_3, \bm r_4) n^{(a)}(\bm r_3, \bm r_4,{-}z)\bigr] \nonumber \\ 
   & \times L^{\text{c}}(\bm r_1, \bm r_2, \bm r_5, \bm r_6,z) V_{\text{p}}^{(b)}(\bm r_5,\bm r_6, z)   . \label{Cab = [vp -kcn]n}
\end{align}
Further we use the connection between the partially interacting propagator $L^{\text{c}}$ and the induced density, given by Eq.\ \eqref{n = Lc vp}, and the symmetry property of $K^{\text{c}}$ expressed in Eqs.\ \eqref{Kc(1234) = Kc(3412)}-\eqref{Kc(1234) = Kc(4321)}. With these ingredients, Eq.\ \eqref{Cab = [vp -kcn]n} is rewritten as 
\begin{align}
   & C_{ab}^{\text{pp}}(z){=}  V_{\text{p}}^{(a)}(\bm r_2,\bm r_1,{-}z) L^{\text{c}}(\bm r_1,\bm r_2,\bm r_3,\bm r_4,z)V_{\text{p}}^{(b)}(\bm r_3,\bm r_4,z) \nonumber\\
    &{-}2n^{(a)}(\bm r_2,\bm r_1,{-}z)K^{\text{p}}(\bm r_1,\bm r_2,\bm r_3,\bm r_4)n^{(b)}(\bm r_3,\bm r_4,z).\label{Cab partial screen}
\end{align}
The first term in Eq.\ \eqref{Cab partial screen} is made up of two partially screened vertices $ V_{\text{p}}^{(a)},  V_{\text{p}}^{(b)}$, connected by the partially interacting propagator $L^{\text{c}}$. We refer to the formulation in Eq.\ \eqref{Cab partial screen} as the `partial screen-partial screen' response function, denoted  $C^\text{pp}_{ab}(z)$.

Eq.\ \eqref{Cab partial screen} represents a further equivalent rewriting of the response function. In close analogy with the screen-screen formulation presented in Sec.\ \ref{subsec: screen screen}, we show that the partial screen-partial screen response function is variational with respect to the density matrix.

\subsubsection{Stationary response functional}
To demonstrate the variationality of the partial screen-partial screen  response, we introduce the response functional 
\begin{widetext}
\begin{align}
     &F_{ab}^{\text{pp}}[\lambda,\lambda\p,z] {=}\! \left[ V_{\ext}^{(a)}(\bm r_1)\delta(\bm r_1{-}\bm r_2) {+} K^{\text{p}}(\bm r_2, \bm r_1, \bm r_5, \bm r_6)\lambda(\bm r_5, \bm r_6) \right] L^{\text{c}}(\bm r_1,\bm r_2,\bm r_3,\bm r_4,z) \nonumber\\
     &\times \left[V_{\ext}^{(b)}(\bm r_3)\delta(\bm r_3{-}\bm r_4) {+} K^{\text{p}}(\bm r_3, \bm r_4, \bm r_7, \bm r_8)\lambda\p(\bm r_7, \bm r_8)\right] - 2\lambda(\bm r_2,\bm r_1) K^{\text{p}}(\bm r_1,\bm r_2,\bm r_3,\bm r_4) \lambda\p(\bm r_3,\bm r_4)  ,\label{Fpp[rho,rho']}
\end{align}  
\end{widetext}
that satisfies the relation  
\begin{equation}
    \label{Cpp = Fpp }
    C_{ab}^{\text{pp}}(z) = F_{ab}^{\text{pp}}[\lambda {=} n^{(a)}({-}z),\lambda\p {=} n^{(b)}(z),z].
\end{equation}
This functional can be obtained replacing into the functional $F_{ab}[\lambda,\lambda\p,z]$ used for $C_{ab}^{\text{ss}}(z)$, Eq.\ \eqref{F[rho,rho']}, $L^{\text{0}}$ with $L^{\text{c}}$ and $K$ with $K^{\text{p}}$. Since the substituted objects share the same properties of  the original ones, we can follow the proof presented in Sec. \ref{stationary proof}, obtaining:
\begin{eqnarray}
 \label{dFpp1 = 0}
    \frac{\delta F_{ab}^{\text{pp}}[\lambda,\lambda\p,z]}{\delta \lambda(\bm r_i, \bm r_j)}\eval_{\substack{\lambda =n^{(a)}({-}z),\\ \lambda\p =n^{(b)}(z)} } &=& 0 ,\\
    \quad
    \frac{\delta F_{ab}^{\text{pp}}[\lambda,\lambda\p,z]}{\delta \lambda\p(\bm r_i, \bm r_j)}\eval_{\substack{\lambda =n^{(a)}(-z),\\ \lambda\p =n^{(b)}(z)} } &=& 0.\label{dFpp2 = 0}
\end{eqnarray}
Eqs.\ \eqref{dFpp1 = 0}-\eqref{dFpp2 = 0} imply that the partial screen-partial screen response function of Eq.\ \eqref{Cab partial screen} is also an extremal point of the functional $F^{\text{pp}}_{ab}[\lambda,\lambda\p,z]$, in both $\lambda$ and $\lambda\p$. In other words, an error of $\delta n^{(b/a)}(\bm r, \bm r\p,z)$ on the density matrix results in an error of order $\left[\delta n^{(b/a)}(\bm r, \bm r\p,\pm z)\right]^2$ in the partial screen-partial screen  response.

\subsubsection{Example of a partial screening approach}

\begingroup
\setlength{\tabcolsep}{12pt}
\renewcommand{\arraystretch}{2}
\begin{table*}[t] 
    \centering
    \begin{tabular}{|c c c|}
    \hline
    \raisebox{0.2\height}{Eq.} & \raisebox{0.2\height}{Expression} & \raisebox{0.2\height}{Diagramatic representation}  \\
    \hline
    \multicolumn{3}{|c|}{\rule{0pt}{0.0cm}}\\
    \eqref{Lc = Lo + LoWLc} & $L^{\mathrm{c}} = L_0 -\frac{1}{2}L_0WL^{\mathrm{c}}$&\includegraphics[keepaspectratio,valign=c,height = 0.6cm]{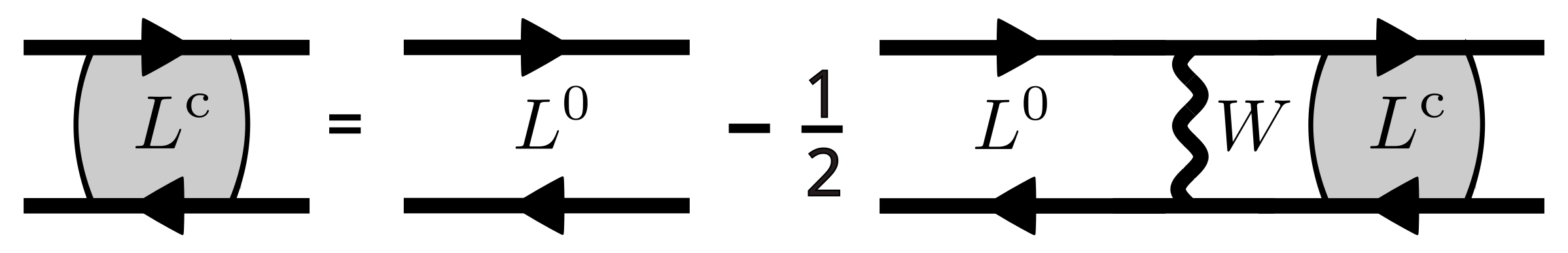}\\[1cm]
    \eqref{L_lad = L_c} & $\chi^{W}(\bm r, \bm r') = L^{\mathrm{c}}(\bm r, \bm r, \bm r',\bm r')$&\includegraphics[keepaspectratio,valign=c,height = 0.6cm]{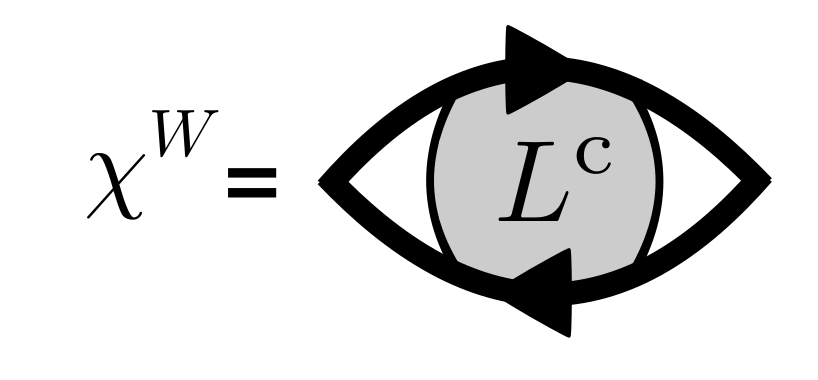} \\[1cm]
     \eqref{bse} & $L = L_0 +L_0vL-\frac{1}{2}L_0WL$ &\includegraphics[keepaspectratio,valign=c,height = 0.6cm]{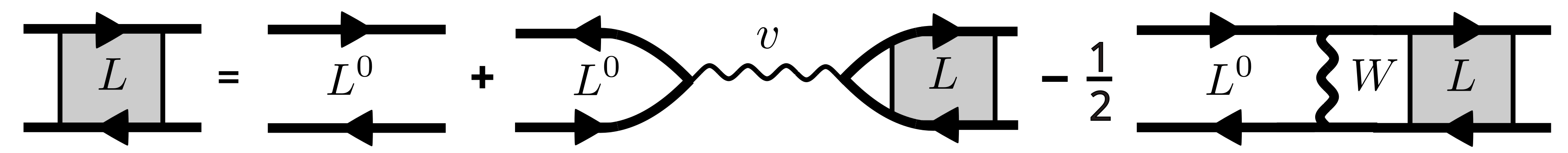} \\[1cm]
    \hline
    \end{tabular}
    \caption{{Diagramatic representation of the the self-consistent equations for the Dyson equation of the interacting electron hole propagators $L^{\mathrm{c}}$, $\chi^W$ and $L$, in the BSE framework where ${\delta V_{\xc}^{\loc}}/{\delta \rho}{=}0$. The bare electron-hole propagator $L^{0}$ is represented as two lines. The unscreened and screened Coulomb interactions, $v$ and $W$ respectively, are represented by thin and thick wavy lines.}}
    \label{table:diagrams L}
\end{table*}
\endgroup

Here we present an example of a particular choice for the separation of the interaction kernel \eqref{k(z) = k1 + k2}. We work in the BSE framework, that corresponds to setting  ${\delta V_{\xc}^{\loc}}/{\delta \rho}{=}0$ in Eq.\ \eqref{fdef}. In this picture, we separate the kernel in Hartree and exchange contribution, selecting for $K^{\text{p}}$ the Hartree kernel i.e.\ 
\begin{align}
    \label{Kp = KH}
    K^{\text{p}}(\bm r,\bm r', \bm r''\!,\bm r''') &= K^{\text{H}}(\bm r,\bm r', \bm r''\!,\bm r''') \nonumber \\
    &= 2  \frac{e^2}{|\bm r - \bm r'' |}\delta(\bm r{-}\bm r')\delta(\bm r''{-}\bm r''')
\end{align}
while the complementary part $K^{\text{c}}$ is the screened-exchange term featuring $W$, i.e.\ 
\begin{align}
    K^{\text{c}}(\bm r,\bm r', \bm r''\!,\bm r''') &=K^{W}(\bm r,\bm r', \bm r''\!,\bm r''') \nonumber \\
    &= - W(\bm r,\bm r') \delta(\bm r{-}\bm r'') \delta(\bm r' {-} \bm r''') \label{Kc = KW}  .
\end{align}
Considering the Dyson equation \eqref{Lc dyson} for the partially interacting electron-hole propagator $L^{\text{c}}$ with the partial kernel Eq.\ \eqref{Kc = KW}, we get 

\begin{align}
    &L^{\text{c}}(\bm r, \bm r', \bm r''\!, \bm r'''\!\!,z) = L^{0}(\bm r, \bm r', \bm r''\!, \bm r'''\!\!,z) \nonumber \\
    &{-}\tfrac{1}{2}L^0(\bm r,\bm r\p,\bm r_1 ,\bm r_2,z)W(\bm r_1,\bm r_2)  L^{\text{c}}(\bm r_1,\bm r_2,\bm r'' ,\bm r'''\!\!,z) . \label{Lc = Lo + LoWLc}
\end{align}
where it is evident that  $L^{\text{c}}$ includes only ladder diagrams. 
Moreover, since the partially screened potential is in the present case local, we just need in $F_{ab}^{\text{pp}}$ the diagonal elements of $L^{\text{c}}$, that we call the ladder-bubble diagram:
\begin{equation}\label{L_lad = L_c}
\chi^W(\bm r, \bm r',z)=L^{\text{c}}(\bm r, \bm r,\bm r', \bm r',z). 
\end{equation}
A diagrammatic representation of the Dyson equation \eqref{Lc = Lo + LoWLc} of $L^{\text{c}}$ and of Eq.\ \eqref{L_lad = L_c} is presented in Table \ref{table:diagrams L}.

The choice of the partial screening Eq.\ \eqref{Kp = KH} bears the important consequence that the partial screened-partial screen response $C^{{\text{pp}}}_{ab}(z)$ in Eq.\ \eqref{Cab partial screen} only depends on the electronic densities, and not on the whole density matrix. Moreover, if we use approximated densities $\rho_\text{ap}^{\text{(b)}}(\bm r,z)$ and $\rho_\text{ap}^{\text{(a)}}(\bm r,{-}z)$ in the variational functional $F_{ab}^{\text{pp}}$, 
 we obtain a variationally approximated expression of $C_{ab}^{\text{pp}}(z)$ as:
\begin{align}
    C_{ab, \text{ap}}^{\text{pp}}(z)&{=}  V_{\text{extH,ap}}^{(a)}(\bm r_1,{-}z) \chi^W(\bm r_1,\bm r_2,z)\
    V_{\text{extH,ap}}^{(b)}(\bm r_2,z) \nonumber\\
    &{-}\rho_{\text{ap}}^{(a)}(\bm r_1,{-}z)\frac{e^2}{|\bm r_1{-} \bm r_2 |}\rho_{\text{ap}}^{(b)}(\bm r_2,z)  ,\label{Cab partial screen ap}
\end{align}
where the approximated partially screened potential is
\begin{align}
\label{defapppot}
V_\text{extH,ap}^{{(b)}}(\bm r,z) 
  &=  V_{\text{ext}}^{{(b)}}(\bm r) {+}\frac{e^2}{|\bm r - \bm r_1|}\rho^{{(b)}}_\text{ap}(\bm r_1, z)   ,
\end{align}
and the analogous expression is used for $V_\text{extH,ap}^{{(a)}}(\bm r,{-}z)$. 
Note that, with this choice of $K^p$, the variational response \eqref{Cab partial screen ap} is similar to the formulation proposed in Ref.\ \cite{stefanucci2024phononlinewidthsscreenedelectronphonon}, cf.\ Eq.\ (6) of the v2 version of the preprint.

We obtain the exact response $C_{ab}^{\text{pp}}(z){=}C_{ab}(z)$ using in Eqs.\ \eqref{Cab partial screen ap} and \eqref{defapppot} the non-approximated densities:
\begin{eqnarray}
    \rho^{{(b)}}(\bm r, z)&=&L(\bm r, \bm r,
   \bm r_1, \bm r_1,z)V_{\text{ext}}^{{(b)}}(\bm r_1),\\
      \rho^{{(a)}}(\bm r, -z)&=&L(\bm r, \bm r,
   \bm r_1, \bm r_1,-z)V_{\text{ext}}^{{(a)}}(\bm r_1),
\end{eqnarray}
where $L(\bm r, \bm r,\bm r', \bm r',z)$ is the full BSE bubble diagram, obtained from the interacting-electron hole propagator $L$. The BSE for $L$ is reported in Table \ref{table:diagrams L}.

By virtue of the variationality of the partial screen-partial screen response Eq.\ \eqref{dFpp1 = 0}-\eqref{dFpp2 = 0}, we have that the error on the response is quadratic in the error on the density: \begin{equation}
    C_{ab, \text{ap}}^{\text{pp}}(z) - C_{ab}(z) = O\left(\left( \rho^{(b/a)}_{\text{ap}}(\pm z) -\rho^{(b/a)}(\pm z) \right)^2\right).
\end{equation}

As we will show in a following paper \cite{usarticolo4}, the approximated scheme given by Eqs.\ \eqref{Cab partial screen ap} and \eqref{defapppot}, is particularly appealing, if we want to devise a practical scheme to evaluate the  BSE corrections to the phonon self-energy on top of a standard static phonon calculation, based on DFT linear response theory \cite{giannozzi1991ab,gonze1997dynamical}. In this case, we can leverage the fact that both the dynamical and BSE effects  produce small relative-corrections to the static DFT electronic densities $\rho^{(b/a)}_{\rm DFT}$ in the range of frequencies corresponding to the phonon scale (typically up to 10{-}100 meV). Thus, in Eqs.\ \eqref{Cab partial screen ap} and \eqref{defapppot}, we can safely use:
\begin{eqnarray}
    \rho^{{(b)}}_\text{ap}(\bm r, z)&=&\rho^{(b)}_{\rm DFT}(\bm r)
    ,\\
      \rho^{{(a)}}_\text{ap}(\bm r,{-}z)&=&\rho^{(a)}_{\rm DFT}(\bm r),
\end{eqnarray}
keeping the resulting error on the phonon self-energy under control, thanks to the variational properties of the $F_{ab}^{\text{pp}}$ functional.

{In Ref.\ \cite{usarticolo4} we will leverage the variationality of the partial screen-partial screen response to effectively approximate the phonon dispersion of graphene including exchange effects. We report here the formal derivation of the partial screen-partial screen formulation in order to focus more on physical insights of our findings in Ref.\ \cite{usarticolo4}.}

\section{Memory effects in the full or partial interaction kernel} \label{sec: retardation}
In the framework used throughout all our manuscript, we have neglected {memory} effects in the electronic interaction, i.e.\ we have worked in the adiabatic approximation Eq.\ \eqref{V Hxc adiabatic}. In this Section, we discuss briefly how to overcome such an approximation. 

To relax the adiabiatic approximation for the exchange and correlation potential, introduced in Eq.\ \eqref{V Hxc adiabatic}, we generalize the nonlocal kernel of the interaction introduced in Eq.\ \eqref{kerneleq} accounting for {memory} effects in the Hartree-exchange-correlation part as 
\begin{align}
\label{kernel retarded}
K(\bm r, \bm r\p,\bm r\pp, \bm r{'''},z)&=2f_{\Hxc}(\bm r,\bm r''\!,z)\delta(\bm r{-}\bm r')\delta(\bm r''{-}\bm r''')\nonumber\\
&{-} W(\bm r,\bm r') \delta(\bm r{-}\bm r'') \delta(\bm r' {-} \bm r''') ,
\end{align}
where $f_{\Hxc}(\bm r,\bm r''\!,z)$ is the retarded Hartree plus exchange-correlation kernel, which is holomorphic for $\Im z{>}0$. A formal expression for  $f_{\Hxc}(\bm r,\bm r''\!,z)$ is presented e.g.\ in Ref.\ \cite{gross1985local}. For the present discussion, we do not discuss a specific expression for $f_{\Hxc}(\bm r,\bm r''\!,z)$ analogous to \eqref{fdef}. We only use the property (see Eq.\  (6) of Ref.\ \cite{gross1985local})
\begin{equation}
    \label{f 12 z= f21 z}
    f_{\Hxc}(\bm r,\bm r''\!,z) {=} f_{\Hxc}(\bm r''\!,\bm r,{-}z) .
\end{equation}
From Eq.\ \eqref{f 12 z= f21 z}, it follows that the retarded kernel of the interaction satisfies \cite{casida1995recent,gross1985local}
\begin{equation}
    \label{K 1234 -z = K 4321 z}
    K(\bm r, \bm r\p,\bm r\pp, \bm r{'''},z) = K(\bm r'''\!\!, \bm r''\!,\bm r', \bm r,{-}z) 
\end{equation}
This property can be leveraged to generalize the formalism developed in Secs.\ \ref{sec:sb ss s*s} and \ref{sec: pp} including memory effects in the Hxc part of the electronic interaction. Specifically, the bare-screen [Eq.\ \eqref{Cab screen bare}], the screen-screen [Eq.\ \eqref{Cab screen screen}], the screen*-screen [Eq.\ \eqref{Cab screen* screen z*}]  and the partial screen-partial screen [Eq.\ \eqref{Cab partial screen}] response can be directly generalized including a frequency dependent kernel, of the form Eq.\ \eqref{kernel retarded}. 
Using the symmetry property Eq.\ \eqref{K 1234 -z = K 4321 z}, it is straightforward to show that the variationality of the screen-screen and of the partial screen-partial screen formulations still hold, while the bare-screen and the screen*-screen formulations are not.

Notice however that the screen*-screen formulation with a retarded kernel does not posses the generalized Fermi Golden Rule structure as in Eq.\ \eqref{fermi golden rule}. In fact, the imaginary part of the screen*-screen response function [Eq.\ \eqref{ImCs*s}] acquires a finite contribution associated with the imaginary part of the retarded kernel.

In conclusion, we mention that the partition of the electronic interaction discussed in Sec.\ \ref{sec: pp} can be performed considering memory in the partial and complementary kernels that sum up to a frequency independent kernel, i.e.\ 
\begin{equation} \label{K = Kp(z) + Kc(z)}
    K(\bm r,\bm r',\bm r''\!,\bm r''') {=} K^{\text{p}}(\bm r,\bm r',\bm r''\!,\bm r'''\!\!,z) {+} K^{\text{c}}(\bm r,\bm r',\bm r''\!,\bm r'''\!\!,z)\:.
\end{equation}
It is straightforward to show that the variationality of the partial screen-partial screen response still holds if the partial retarded kernels satisfy the properties
\begin{align}
    %\label{Kp 1234 -z = Kp 4321 z}
    K^{\text{p}}(\bm r, \bm r\p,\bm r\pp, \bm r{'''},z) &= K^{\text{p}}(\bm r'''\!\!, \bm r''\!,\bm r', \bm r,{-}z), \\
    K^{\text{c}}(\bm r, \bm r\p,\bm r\pp, \bm r{'''},z) &= K^{\text{c}}(\bm r'''\!\!, \bm r''\!,\bm r', \bm r,{-}z)\:.
\end{align}
A separation of the form Eq.\ \eqref{K = Kp(z) + Kc(z)} can be useful in instances where the memory effects are included in the individual parts of the electronic screening, but cancel out when summed in the fully interacting kernel.

%% +++++ AGGIUNGERE ++++++ %%%
%% proprieta K* = K commento francesco
%%%%%%%%%%%%%%%%%%%%%%%%%%%%%

\section{Application: optical conductivity of graphene} \label{sec: application to graphene}
In this Section, we apply the linear response formalism developed in the previous sections to the calculation of optical conductivity of graphene. As detailed in Ref.\ \cite{albertoTB}, the diagonal element of the optical conductivity tensor in the direction of the unitary vector $\hat{\bm q}{=} \bm q/q$
%of a 2D material, with the rectangular approximation of the nonperiodic part of the orbitals, 
can be computed as 
\begin{equation} \label{sigma 2d alberto}
    \sigma_{\hat{\bm q}} (z)= \lim_{q\rightarrow 0}\frac{iz\bar{\chi}(\bm{q},z)}{q^2} \ .
\end{equation}
%\FM{se mettiamo il barrato l'equazione e' valida anche in 3D. Giusto? Quindi ho tolto i riferimenti al caso 2D. Solo nelle di chi non barrata l'equazione e' valida solo in 2D.}
In Eq.\ \eqref{sigma 2d alberto}, $\bm q$ is the momentum of the external scalar potential, and $\bar{\chi}$ is the macroscopic density-density response computed in the condition of null macroscopic electric field (this condition is indicated with a bar over the symbol representing the physical quantity, to be consistent with the notations introduced in Refs.\ \cite{albertoTB,PhysRevLett.129.185902,PhysRevB.88.174106,PhysRevB.107.094308,PhysRevB.110.094306}). In the small $\bm q$ regime considered in this Section, the null macroscopic field condition is imposed by removing the macroscopic component of the Coulomb kernel in the Hartree term of Eq.\ \eqref{kerneleq}, $\bar{v}_{\bm{G}=0}(\bm{q}) {=} 0$, where $v_{\bm{G}}(\bm{q})$ is the Fourier transform of the Coulomb potential and $\bm{G}$ are the reciprocal lattice vectors. 
%\FM{Domanda: questo lo facciamo solo sul termine di Hartree o anche su quello di scambio con W? Cosa succede se mettiamo o non mettiamo mil termine macroscopico nello scambio? Il termine long range nello scambio in un isolante 3D e' quello vhe va regolarizzato. Nel metallo copn la W statica e' regolarizzato dallo scrining ma noin nel caso dinamico. Forse la questione e' complkicata...Qui comunque espliciterei in modo esplicito quele e' la nostra definizione di barramento ov vero se la mettiamo solo su Hartree o anche di W}

In the following, we present the formulas for $\chi$ and $\bar \chi$ and we show the results for $\sigma$, which can be directly linked to an experimental measure, e.g.\ optical absorption. 

%Since the density-density is a response to potentials in the form of Eq. \eqref{Vext = Vext compl}, the functionals of Eqs. \eqref{F sb} and \eqref{F[rho,rho']} will be indicated in the following as $\mathcal{F}_{\bm q}[\lambda,\lambda\p,z]$ and $F_{\bm q}[\lambda,\lambda\p,z]$. The numerical details of the implementation are given in Appendix \ref{sec-app:TB implementation}.

\subsection{Density-density response in periodic systems} 
\label{subsec: variationality proof}
The eigenstates of $\hat H^{(0)}$ are Bloch states $|\psi^{(0)}_{m\bm k}\rangle$ of quasimomentum $\bf k$, normalized on the Born–von Karman supercell of volume $N\Omega$, containing $N$ unit cells.
If $V_\ext^{(b)}(\bm r){=}V_\ext^{(\bm q)}(\bm r){=}e^{i\bm q {\cdot} \bm r}$, we can write $\hat n^{(\bm q)}$ and $\hat V_{\scf}^{(\bm q)}$  in the basis of the unperturbed states as:
\begin{align}
&\langle  \psi^{(0)}_{n\bm k'}| \hat n^{(\bm q)}(z)|\psi^{(0)}_{m\bm k}\rangle {=} \delta_{\bm k',\bm k+ \bm q}
\langle  \psi^{(0)}_{n\bm k + \bm q}| \hat n^{(\bm q)}(z)|\psi^{(0)}_{m\bm k}\rangle, \\
&\langle  \psi^{(0)}_{n\bm k'}| \hat V_{\scf}^{(\bm q)}(z)|\psi^{(0)}_{m\bm k}\rangle {=} \delta_{\bm k',\bm k+ \bm q}
\langle  \psi^{(0)}_{n\bm k +\bm q}| \hat V_{\scf}^{(\bm q)}(z)|\psi^{(0)}_{m\bm k}\rangle.\! 
\end{align}
In this representation,  Eqs.\ \eqref{n(r) = L0V  (r)} and \eqref{L0(r)} express the relation between the nonzero matrix elements of $\hat n^{(\bm q)}$ and $\hat V_{\scf}^{(\bm q)}$ as
\begin{eqnarray}\label{app: sternheimer}
2\langle  \psi^{(0)}_{n\bm k + \bm q}| \hat n^{(\bm q)}(z)|\psi^{(0)}_{m\bm k}\rangle&=&
L^{0}_{n\bm k {+} \bm q, m\bm k}(z)
\nonumber \\ 
& \times &
\langle  \psi^{(0)}_{n\bm k +\bm q}| \hat V_{\scf}^{(\bm q)}(z)|\psi^{(0)}_{m\bm k}\rangle \label{nqfromvscf}
,
\end{eqnarray}
where
\begin{align}
L^{0}_{n\bm k {+} \bm q, m\bm k}(z)
   = 2 \frac{f_{m\bm k}\unpert - f_{n\bm k +\bm q}\unpert}{\hbar z - (\varepsilon\unpert_{n\bm k +\bm q} -\varepsilon\unpert_{m\bm k})}.
\end{align}
Eqs.\ \eqref{V_scf(r) = V + Kn} and \eqref{kerneleq} become:
\begin{eqnarray}
&& \langle  \psi^{(0)}_{n\bm k +\bm q}| \hat V_{\scf}^{(\bm q)}(z)|\psi^{(0)}_{m\bm k}\rangle= 
\langle  \psi^{(0)}_{n\bm k +\bm q}| \hat V_{\ext}^{(\bm q)}|\psi^{(0)}_{m\bm k}\rangle 
\nonumber \\&&  {+}\frac{1}{N}\sum_{\bm k\p sl}
 K^{n\bm k {+} \bm q, m\bm k}_{s\bm k' {+} \bm q, l\bm k'}
\langle  \psi^{(0)}_{s\bm k' + \bm q}| \hat n^{(\bm q)}(z)|\psi^{(0)}_{l\bm k'}\rangle
, \label{app: V_scf = V + Kn}
\end{eqnarray}
where, in the BSE framework
\begin{eqnarray}\label{kernel sx}
K^{n\bm k {+} \bm q, m\bm k}_{s\bm k' {+} \bm q, l\bm k'}
&=& N\!\!\int\limits_{N\Omega}\!\! \dd{\bm r} \left[
\psi^{(0)}_{n\bm k+\bm q}(\bm r)^*\psi^{(0)}_{m\bm k}(\bm r)2 v(\bm r-\bm r_1)\right.
\nonumber \\
&\times& \psi^{(0)}_{s\bm k\p+\bm q}(\bm r_1) \psi^{(0)}_{l\bm k\p}(\bm r_1)^*-
\psi^{(0)}_{n\bm k+\bm q}(\bm r)^*\psi^{(0)}_{m\bm k}(\bm r_1) \nonumber \\
&\times& \left.W(\bm r, \bm r_1)\psi^{(0)}_{s\bm k\p+\bm q}(\bm r) \psi^{(0)}_{l\bm k\p}(\bm r_1)^*\right],
\end{eqnarray}
where, as usual, the integral over ${\bm r}$ runs over the $N$ cells of the supercell and the implicit one over $\bm r_1$ on the full crystal.

Usually, the most difficult (i.e.\ the most computationally time
consuming) quantities to be evaluated with accuracy are the
self-consistent electronic density matrix and potential,
since they need to be determined via the self-consistent set of Eqs.~\eqref{nqfromvscf}, \eqref{app: V_scf = V + Kn}.
On the contrary, $L^{0}_{n\bm k {+} \bm q, m\bm k}(z)$ has a very simple explicit form, that only
requires the knowledge of energies and wavefunctions of
the unperturbed Hamiltonian $\hat H^{(0)}$. In the evaluation of $\bar{\chi}$, it is convenient to use an approximated density matrix:
\begin{align}\label{eq:substi}
\langle  \psi^{(0)}_{n\bm k + \bm q}| \hat n^{(\bm q)}(z)|\psi^{(0)}_{m\bm k}\rangle_{\text{ap}}.
\end{align}
From this approximated density matrix, we define a corresponding approximated self-consistent potential as:
\begin{eqnarray}
&& \langle  \psi^{(0)}_{n\bm k +\bm q}| \hat V_{\scf}^{(\bm q)}(z)|\psi^{(0)}_{m\bm k}\rangle_{\text{ap}}= 
\langle  \psi^{(0)}_{n\bm k +\bm q}| \hat V_{\ext}^{(\bm q)}|\psi^{(0)}_{m\bm k}\rangle 
\nonumber \\&&  {+}\frac{1}{N}\sum_{\bm k\p sl}
 K^{n\bm k {+} \bm q, m\bm k}_{s\bm k' {+} \bm q, l\bm k'}
\langle  \psi^{(0)}_{s\bm k' + \bm q}| \hat n^{(\bm q)}(z)|\psi^{(0)}_{l\bm k'}\rangle_{\text{ap}}
. \label{appoximated: V_scf = V + Kn}
\end{eqnarray}
Notice that, in general, for the approximated quantities Eq.~\eqref{nqfromvscf} does not hold.

Within the functional formulation defined in Section \ref{sec:sb ss s*s}, we obtain the approximated expressions of ${\chi}$ by using {Eqs.~\eqref{Csb = Fab}, \eqref{Css = Fab}, \eqref{Cs*s = Fab}}, where in place of $\lambda$ and $\lambda'$ we use the approximated density matrix (Eq.\ \eqref{eq:substi}), obtaining: 
\begin{widetext}
\begin{eqnarray}
\label{chi sb in kspace}
{\chi}^{\text{bs}}_{\text{ap}}(\bm{q},z)
=  \frac{1}{N}\sum_{\bm k n  m }
\langle  \psi^{(0)}_{m\bm k}| \hat V_{\ext}^{(-\bm q)}|\psi^{(0)}_{n\bm k +\bm q}\rangle
L^{0}_{n\bm k {+} \bm q, m\bm k}(z)
\langle  \psi^{(0)}_{n\bm k +\bm q}| \hat V_{\scf}^{(\bm q)}(z)|\psi^{(0)}_{m\bm k}\rangle_{\text{ap}}.
\end{eqnarray} 
Instead, the screen-screen and the screen*-screen density-density response read 
\begin{eqnarray}
    \label{chi ss in k space}
{\chi}^{\text{ss}}_{\text{ap}}(\bm{q},z) &=& \frac{1}{N}\sum_{\bm k n  m }
\langle  \psi^{(0)}_{m\bm k}| \hat V_{\scf}^{(-\bm q)}({-}z)|\psi^{(0)}_{n\bm k +\bm q}\rangle_{\text{ap}}
L^{0}_{n\bm k {+} \bm q, m\bm k}(z)
\langle  \psi^{(0)}_{n\bm k +\bm q}| \hat V_{\scf}^{(\bm q)}(z)|\psi^{(0)}_{m\bm k}\rangle_{\text{ap}} \nonumber  \\
&&-\frac{2}{N^2}
\sum_{\substack{\bm k   nm \\ \bm k \p l s}}
\langle \psi^{(0)}_{m\bm k}|\hat n^{(-\bm q)}({-}z)
|\psi^{(0)}_{n\bm k+\bm q}\rangle_{\text{ap}}
 K^{n\bm k {+} \bm q, m\bm k}_{s\bm k' {+} \bm q, l\bm k'}
\langle  \psi^{(0)}_{s\bm k' + \bm q}| \hat n^{(\bm q)}(z)|\psi^{(0)}_{l\bm k'}\rangle_{\text{ap}}, \\
{\chi}^{\text{s*s}}_{\text{ap}}(\bm{q},z) &=& \frac{1}{N}\sum_{\bm k n  m }
\langle  \psi^{(0)}_{m\bm k}| \hat V_{\scf}^{(-\bm q)}({-}z^*)|\psi^{(0)}_{n\bm k +\bm q}\rangle_{\text{ap}}
L^{0}_{n\bm k {+} \bm q, m\bm k}(z)
\langle  \psi^{(0)}_{n\bm k +\bm q}| \hat V_{\scf}^{(\bm q)}(z)|\psi^{(0)}_{m\bm k}\rangle_{\text{ap}} \nonumber  \\
&&-\frac{2}{N^2}
\sum_{\substack{\bm k   nm \\ \bm k \p l s}}
\langle \psi^{(0)}_{m\bm k}|\hat n^{(-\bm q)}({-}z^*)
|\psi^{(0)}_{n\bm k+\bm q}\rangle_{\text{ap}}
 K^{n\bm k {+} \bm q, m\bm k}_{s\bm k' {+} \bm q, l\bm k'}
\langle  \psi^{(0)}_{s\bm k' + \bm q}| \hat n^{(\bm q)}(z)|\psi^{(0)}_{l\bm k'}\rangle_{\text{ap}}. \label{chi s*s in k space}
\end{eqnarray} 
\end{widetext}

Notice that in Eqs.\ \eqref{chi ss in k space},\eqref{chi s*s in k space}, the first matrix elements can be obtained from the relations:
\begin{align}
\langle  \psi^{(0)}_{m\bm k}| \hat V_{\scf}^{(-\bm q)}({-}z)|\psi^{(0)}_{n\bm k +\bm q}\rangle_{\text{ap}}
&=
\langle  \psi^{(0)}_{n\bm k +\bm q}| \hat V_{\scf}^{(\bm q)}(z^*)|\psi^{(0)}_{m\bm k}\rangle_{\text{ap}}^*,\label{eq:-qz*1}\\
\langle  \psi^{(0)}_{m\bm k}| \hat V_{\scf}^{(-\bm q)}({-}z^*)|\psi^{(0)}_{n\bm k +\bm q}\rangle_{\text{ap}}
&=
\langle  \psi^{(0)}_{n\bm k +\bm q}| \hat V_{\scf}^{(\bm q)}(z)|\psi^{(0)}_{m\bm k}\rangle_{\text{ap}}^*,\label{eq:-qz*2}
\end{align} 
and that analogous relations holds for the matrix elements of the density matrix.
% \FMau{\\Qui al posto del paragrafo precedente possiamo mettere.\\
% Notice that in Eqs.\ \eqref{chi ss in k space},\eqref{chi s*s in k space}, the first matrix elements can be obtained from the relations:
% \begin{align}
% \langle  \psi^{(0)}_{m\bm k}| \hat V_{\scf}^{(-\bm q)}({-}z)|\psi^{(0)}_{n\bm k +\bm q}\rangle_{\text{ap}}^*
% =
% \langle  \psi^{(0)}_{n\bm k +\bm q}| \hat V_{\scf}^{(\bm q)}(z^*)|\psi^{(0)}_{m\bm k}\rangle_{\text{ap}}^*,\label{eq:-qz*1}\\
% \langle  \psi^{(0)}_{m\bm k}| \hat V_{\scf}^{(-\bm q)}(-z^*)|\psi^{(0)}_{n\bm k +\bm q}\rangle_{\text{ap}}
% =
% \langle  \psi^{(0)}_{n\bm k +\bm q}| \hat V_{\scf}^{(\bm q)}(z)|\psi^{(0)}_{m\bm k}\rangle_{\text{ap}}^*,\label{eq:-qz*2}
% \end{align} 
% and that analogous relations holds for the matrix elements of the density matrix. ...Equazioni da verificare... Se le equazioni qui sopra sono corrette, i conti numerici fatti in precedenza sono immodificati. Ioltre non mi pare utile in nessuna perte del lavoro invocare, commentere o usare il time reversal...nel caso propongop di non menzionare mai il time-reversal.
% }

% As far as all the quantities appearing in their expressions are evaluated
% exactly
If no approximation are used, the three formulations provide the exact result ${\chi}(\bm{q},z)$. Instead if $\langle  \psi^{(0)}_{n\bm k + \bm q}| \hat n^{(\bm q)}(z)|\psi^{(0)}_{m\bm k}\rangle_{\text{ap}}\ne \langle  \psi^{(0)}_{n\bm k + \bm q}| \hat n^{(\bm q)}(z)|\psi^{(0)}_{m\bm k}\rangle$, the three formulation will provide three distinct estimations of ${\chi}(\bm{q},z)$. Since the screen-screen formulation is variational with respect to the density matrix [Eq.\ \eqref{dF/drho}], while the 
bare-screen and screen*-screen formulations are not [Eqs.~\eqref{dbsdn} and \eqref{dF/drho' s*s}], we have that:
%\begin{subequations} 
    \begin{align}
   {\chi}^{\text{bs}}_{\text{ap}}(\bm{q},z){-}{\chi}(\bm{q},z)&=O\left(\hat n^{(\bm q)}_{\text{ap}}(z){-}\hat n^{(\bm q)}(z)\right), \label{expected errorbs}\\
    {\chi}^{\text{ss}}_{\text{ap}}(\bm{q},z){-}{\chi}(\bm{q},z)&=O\left(\left(\hat n^{(\bm q)}_{\text{ap}}(z){-}\hat n^{(\bm q)}(z)\right)^2\right), \label{expected errorss}\\
   {\chi}^{\text{s*s}}_{\text{ap}}(\bm{q},z){-}{\chi}(\bm{q},z)&=O\left(\hat n^{(\bm q)}_{\text{ap}}(z){-}\hat n^{(\bm q)}(z)\right).  \label{expected errors*s} 
    \end{align}
%\end{subequations}
Namely, for small deviations of the approximated density matrix from the exact one, the screen-screen formulation provides the best estimation of the exact values, as we demonstrate numerically in the next Sections. 

\subsection{Electron-electron interaction in the density density response of graphene} 
\label{sec:elelint}
To account for the effects of the electron-electron interaction in the response of graphene, we follow the approach thoroughly described in Ref.\ \cite{albertoTB}. We include the interaction in $\hat H^{(0)}$ by the static-screened exchange approximation, and in the density-density response by solving the BSE for $\bar{\chi}(\bm{q},z)$, with $\hbar z{=}\hbar \omega {+} i \eta$. In both calculations, we use the same static $W(\bm r, \bm r')$. We work within the $\pi$-orbitals subspace using a tight-binding model. We obtain the optical conductivity $\sigma(z)$ from Eq.\ \eqref{sigma 2d alberto} using a finite but small value of $|\bm q|$. Due to the symmetries of graphene, the conductivity is independent of the direction $\hat {\bm q}$ and we can drop from $\sigma$ such subscript.  To obtain the required $\bar {\chi}(\bm{q},z)$, we eliminate the long-range component of the Hartree kernel. This is done replacing in all the Eqs.\ of the previous subsection $\ K^{n\bm k {+} \bm q, m\bm k}_{s\bm k' {+} \bm q, l\bm k'}$ 
by $\bar K^{n\bm k {+} \bm q, m\bm k}_{s\bm k' {+} \bm q, l\bm k'}$, as detailed in Appendix \ref{sec-app:TB implementation}.

In the present work,  the self-consistent system of Eqs.~\eqref{app: sternheimer} and \eqref{app: V_scf = V + Kn} is solved using an iterative procedure. Namely, we call $\langle  \psi^{(0)}_{n\bm k + \bm q}| \hat n^{(\bm q)}(z)|\psi^{(0)}_{m\bm k}\rangle_\tau$ the density matrix at the iteration $\tau$. By inserting this density matrix in Eq.~\eqref{app: V_scf = V + Kn} together with $\bar K^{n\bm k {+} \bm q, m\bm k}_{s\bm k' {+} \bm q, l\bm k'}$, we obtain the self-consistent potential $\langle  \psi^{(0)}_{n\bm k + \bm q}| \hat V^{(\bm q)}_{\scf}(z)|\psi^{(0)}_{m\bm k}\rangle_\tau$ at the same iteration $\tau$. At the first iteration ($\tau{=}0$), we use:
\begin{eqnarray}\label{nqfromvscfstart}
2\langle  \psi^{(0)}_{n\bm k + \bm q}| \hat n^{(\bm q)}(z)|\psi^{(0)}_{m\bm k}\rangle_0&=&
L^{0}_{n\bm k {+} \bm q, m\bm k}(z)
\nonumber \\ 
& \times &
\langle  \psi^{(0)}_{n\bm k +\bm q}| \hat V_{\ext}^{(\bm q)}|\psi^{(0)}_{m\bm k}\rangle.
\end{eqnarray} %%% Gio === Vext non dipende da z, cambiato %%%
The density matrix at the iteration $\tau{+}1$ is obtained by  summing the density matrix at the step $\tau$, with a weight $(1{-}x)$, with that obtained inserting the self-consistent potential at the iteration $\tau$ in Eq.~\eqref{app: sternheimer}, with a weight $x$. Typical values of $x{\simeq}0.2$ assures the convergence of the iterative solution to its fixed point.
In Appendix \ref{sec-app:TB implementation}, we summarize other details of the calculation. 

\begin{figure}[t] 
    \centering
    \includegraphics[width = .9\columnwidth, keepaspectratio]{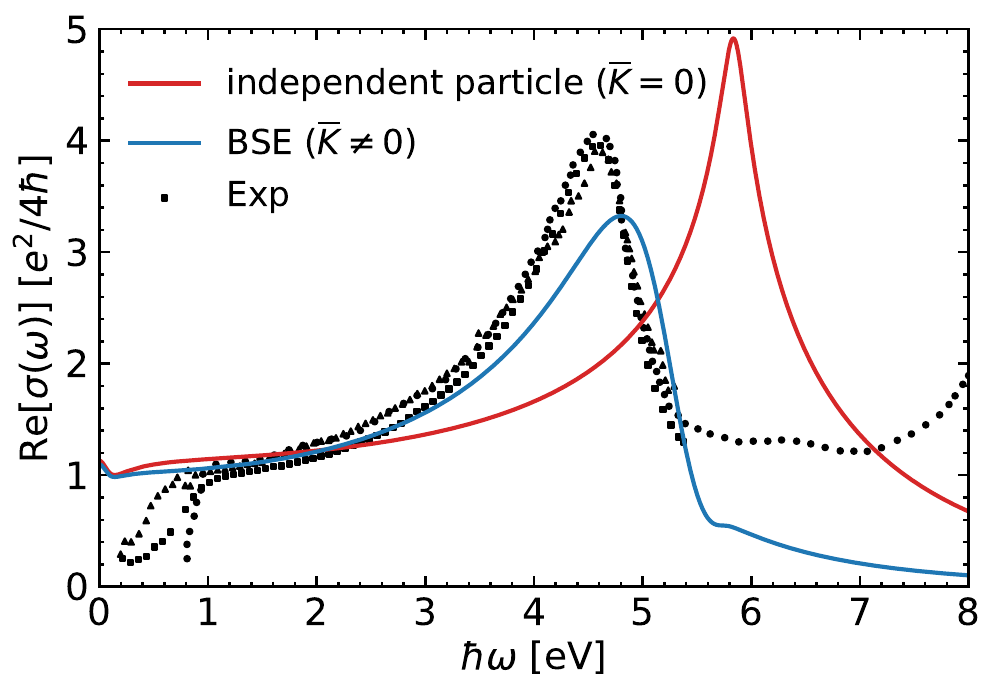}
    \caption{Optical conductivity of freestanding graphene obtained with the independent particle approximation (red), i.e.\ by setting the kernel $\bar K=0$, and by solving the BSE (blue), compared within experimental results (black symbols).
    Black triangles, squares and dots are experimental data obtained from Refs.~\cite{mak2012optical}, \cite{chang2014extracting} and \cite{li2016broadband} respectively.}
\label{fig:sigmavsexp}
\end{figure}

In Fig.~\ref{fig:sigmavsexp}, we show the remarkable agreement between the measured optical conductivity and that predicted by our BSE approach, for $\hbar \omega{<} 5.5$ eV. The measured rise for $\hbar \omega{>} 7$ eV and the plateau between $5.5$ and $7$ eV are missing in the theoretical results, since they are associated to transitions involving  $\sigma$-bands, neglected in our approach.
In Fig.~\ref{fig:sigmavsexp}, we also show the large impact of the excitonic interaction, i.e.\ the ladder corrections with $W$ (cf.\ Table \ref{table:diagrams}). Indeed, if we neglect the  screening of the vertices, by setting to zero  $\bar K^{n\bm k {+} \bm q, m\bm k}_{s\bm k' {+} \bm q, l\bm k'}$ in the calculation, we overestimate the position of the $\pi{-}\pi^*$ plasmon peak (at ${\sim}4.6$ eV in experiments). Moreover, we do not reproduce the expected asymmetric shape of the peak, as proven in Refs.~\cite{Yang2009,Yang2011}.

\subsection{Exploiting variationality}

In the following, we consider two exemplary kinds of errors in the evaluation of the electronic density matrix. The first example regards the error that naturally arises when performing the self-consistent cycle described in Sec.\ \ref{sec:elelint} in terms of $\tau$. The second instead regards the case where the exact electronic density matrix vertex is computed for some given frequency $\omega{=}\omega_0$, and then such density matrix is used as an approximation to determine the response function across the entire frequency range.

\subsubsection{Convergence of the optical conductivity within the iterative approach} \label{subsubsec: convergence study}

\begin{figure}[t]
    \centering
    \includegraphics[width = \columnwidth, keepaspectratio]{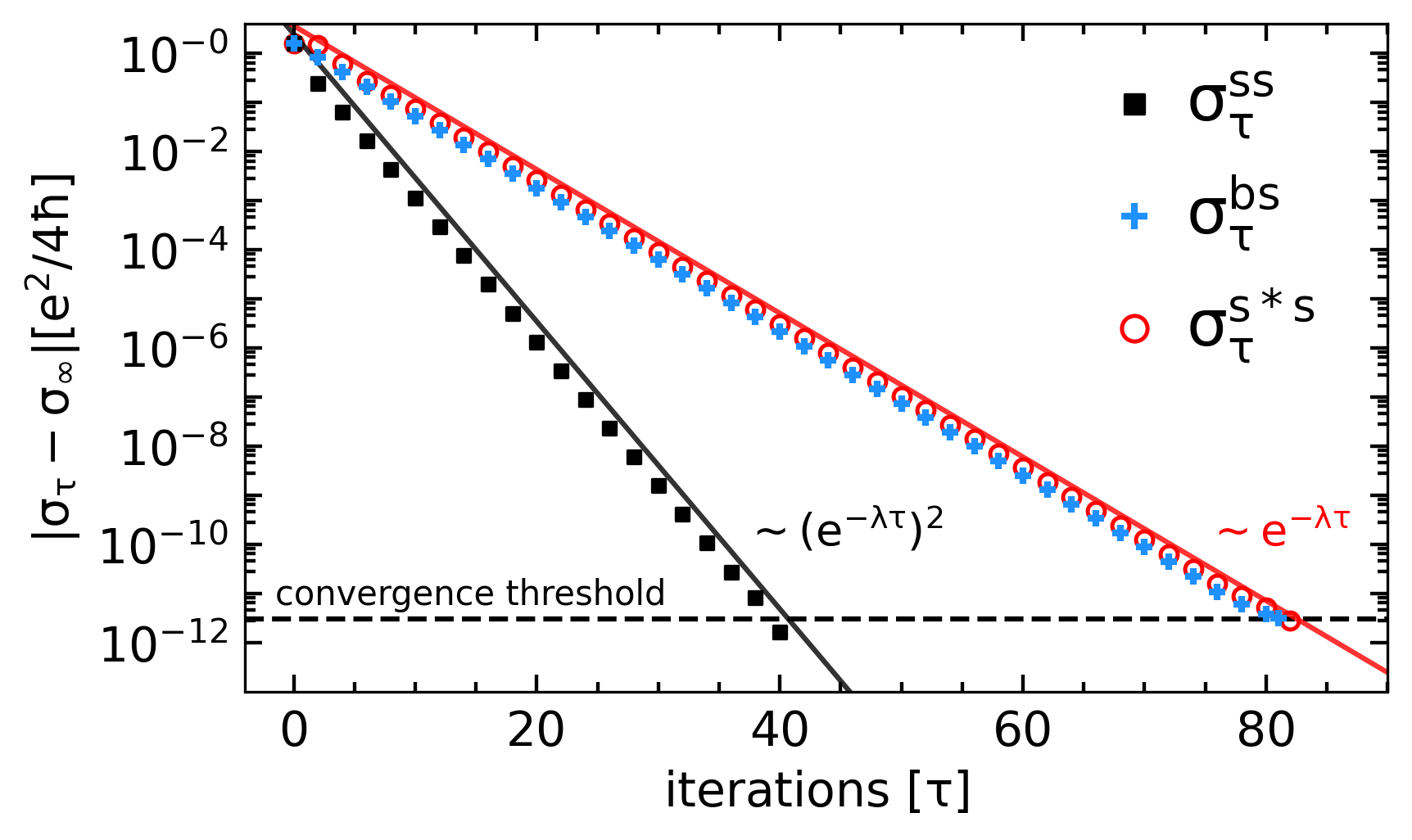}
    \caption{Convergence behavior of the conductivity $\sigma(z)$ at ($\hbar\omega {=} 4.8$ eV, $\hbar\eta {=} 0.1$ eV) as a function of the iteration step $\tau$ of the self-consistent scheme explained in Sec.\ \ref{sec:elelint}, for the three different formulations described in Sec.\ \ref{sec:sb ss s*s} (screen-screen in red, bare-screen in blue, screen*-screen in green). Scatter points indicate the error  $|\sigma_{\tau}{-}\sigma_{\infty}|$. The cycle is terminated when $|\sigma_{\tau}{-}\sigma_{\infty}|$ is below a given convergence threshold (black dashed line). Lines indicate linear fits of the convergence trend, that in semi-log scale indicate exponential convergence. The screen-screen response convergences twice as fast with respect to the other formulations, thus proving the variationality of the screen-screen formulation.}
    \label{fig:sigma fixedw
    }   
    \label{fig:variationality proof}
\end{figure}
We investigate the convergence of the electronic density matrix within the iteration cycle described in Sec.\ \ref{sec:elelint}. Referring to the notation of the previous section, we now perform the following identification
\begin{align}
\langle  \psi^{(0)}_{n\bm k + \bm q}| \hat n^{(\bm q)}(z)|\psi^{(0)}_{m\bm k}\rangle_{\text{ap}} = \langle  \psi^{(0)}_{n\bm k + \bm q}| \hat n^{(\bm q)}(z)|\psi^{(0)}_{m\bm k}\rangle_{\tau},
\end{align}
and the same for all the other quantities for which `ap'$\rightarrow$`$\tau$'. 

It is clear that the scaling properties of $\bar \chi(z)$ with respect to the density are reflected on $\sigma(z)$ via Eq. \eqref{sigma 2d alberto}. Henceforth, we now show numerical results proving Eqs.\ \eqref{expected errorbs},\ \eqref{expected errorss} and \ \eqref{expected errors*s},  but studying the optical conductivity.
We introduce the counterparts of the density-density responses obtained via Eq.\ \eqref{sigma 2d alberto} as $\sigma^{\text{bs}}_{\tau}(z)$, $\sigma^{\text{ss}}_{\tau}(z)$ and $\sigma^{\text{s*s}}_{\tau}(z)$, with obvious meaning of the symbols. We verified numerically that: 
\begin{equation} 
\sigma^{\text{bs}}_{\infty}(z)=\sigma^{\text{ss}}_{\infty}(z)=\sigma^{\text{s*s}}_{\infty}(z)=\sigma(z).
\end{equation}
For each formulation, we report the values of the error as a function of the iteration step in Fig. \ref{fig:variationality proof}, for the frequency $\hbar \omega{=}4.8$ eV. In the semi-logarithmic scale of Fig.\ \ref{fig:variationality proof}, all the formulations converge linearly with respect to the iterations number, meaning that $\sigma_{\tau}(z)$ converges exponentially with the number of iterations $\tau$. Also, we notice that in semi-logarithmic scale $\sigma^{\text{ss}}_{\tau}$ converges twice as fast with respect to both $\sigma^{\text{bs}}_{\tau}$ and $\sigma^{\text{s*s}}_{\tau}$, confirming the advantageous variational property of the screen-screen formulation.

\subsubsection{Fixed frequency approximation of optical conductivity}
\label{subsec: variationality exploit}

In this section we treat the case where the (numerically) exact electronic density matrices known for some given frequency $\omega_0$ (namely, for $z_0{=}\omega_0{+}i\eta$ and $z_0^*{=}\omega_0{-}i\eta$) and then used to determine the response function across the entire frequency range. Such an approximation clearly comes with great numerical advantage, in terms of time saved avoiding repeating the self consistent routine for every frequency. This is often the case of phonon calculations, where the electron-phonon vertexes are known only in the static case ($z_0{=}z_0^*{=}0$). 

In this case, we pose:
\begin{align}
    \langle  \psi^{(0)}_{n\bm k + \bm q}| \hat n^{(\bm q)}(z)|\psi^{(0)}_{m\bm k}\rangle_{\text{ap}}&=\langle  \psi^{(0)}_{n\bm k + \bm q}| \hat n^{(\bm q)}(z_0)|\psi^{(0)}_{m\bm k}\rangle, \label{n(z) = n(z0)}\\
\langle  \psi^{(0)}_{n\bm k + \bm q}| \hat V_{\scf}^{(\bm q)}(z)|\psi^{(0)}_{m\bm k}\rangle_{\text{ap}}&=\langle  \psi^{(0)}_{n\bm k + \bm q}| \hat V_{\scf}^{(\bm q)}(z_0)|\psi^{(0)}_{m\bm k}\rangle \label{V(z) = V(z0)}
\end{align}
and similarly for all the $z^*$ quantities. Using the definitions in Eqs.\ \eqref{n(z) = n(z0)} and \eqref{V(z) = V(z0)}, we get
\begin{equation}
    \bar{\chi}^{\text{bs}}_{\text{ap}}(\bm{q},z_0)=\bar{\chi}^{\text{ss}}_{\text{ap}}(\bm{q},z_0)=\bar{\chi}^{\text{s*s}}_{\text{ap}}(\bm{q},z_0)=\bar{\chi}(\bm{q},z_0)
\end{equation}
and the second summations in Eqs.~\eqref{chi ss in k space} and \eqref{chi s*s in k space} do not depend anymore on $z$. Therefore, we obtain that:
\begin{widetext}
\begin{eqnarray}
\label{chi sb in kspacez0}
\bar{\chi}^{\text{bs}}_{z_0}(\bm{q},z)
&=& \bar{\chi}(\bm{q},z_0)+ \frac{1}{N}\sum_{\bm k n  m }
\langle  \psi^{(0)}_{m\bm k}| \hat V_{\ext}^{(-\bm q)}|\psi^{(0)}_{n\bm k +\bm q}\rangle
\Delta L^{0}_{n\bm k {+} \bm q, m\bm k}(z,z_0)
\langle  \psi^{(0)}_{n\bm k +\bm q}| \hat V_{\scf}^{(\bm q)}(z_0)|\psi^{(0)}_{m\bm k}\rangle,\label{diffbs}\\
    \label{chi ss in k spacez0}
\bar{\chi}^{\text{ss}}_{z_0}(\bm{q},z) &=& \bar{\chi}(\bm{q},z_0)+\frac{1}{N}\sum_{\bm k n  m }
\langle  \psi^{(0)}_{m\bm k}| \hat V_{\scf}^{(-\bm q)}({-}z_0)|\psi^{(0)}_{n\bm k +\bm q}\rangle
\Delta L^{0}_{n\bm k {+} \bm q, m\bm k}(z,z_0)
\langle  \psi^{(0)}_{n\bm k +\bm q}| \hat V_{\scf}^{(\bm q)}(z_0)|\psi^{(0)}_{m\bm k}\rangle, \label{diffss} \\
\bar{\chi}^{\text{s*s}}_{z_0}(\bm{q},z) &=& \bar{\chi}(\bm{q},z_0)+ \frac{1}{N}\sum_{\bm k n  m }
\langle  \psi^{(0)}_{m\bm k}| \hat V_{\scf}^{(-\bm q)}({-}z_0^*)|\psi^{(0)}_{n\bm k +\bm q}\rangle
\Delta L^{0}_{n\bm k {+} \bm q, m\bm k}(z,z_0)
\langle  \psi^{(0)}_{n\bm k +\bm q}| \hat V_{\scf}^{(\bm q)}(z_0)|\psi^{(0)}_{m\bm k}\rangle, 
 \label{chi s*s in k spacez0}
\end{eqnarray} 
\end{widetext}
where, for notation clarity, we replaced the ``$\text{ap}$'' subscript in the approximated $\bar \chi$ with ``$z_0$'' and
\begin{equation}
\Delta L^0_{n\bm k {+} \bm q, m\bm k}(z,z_0)=L^0_{n\bm k {+} \bm q, m\bm k}(z)-L^0_{n\bm k {+} \bm q, m\bm k}(z_0).
\end{equation}
Due to the different variational properties of the three approaches, the error of the screen-screen formulation is predicted to be the smallest for $z$ close to $z_0$, and:
\begin{eqnarray}
\bar{\chi}^{\text{bs}}_{z_0}(\bm{q},z)-\bar{\chi}(\bm{q},z)&=&O\left(z-z_0\right),\\
\bar{\chi}^{\text{ss}}_{z_0}(\bm{q},z)-\bar{\chi}(\bm{q},z)&=&O\left((z-z_0)^2\right), \label{errorss}\\
\bar{\chi}^{\text{s*s}}_{z_0}(\bm{q},z)-\bar{\chi}(\bm{q},z)&=&O\left(z-z_0\right).
\end{eqnarray} 
In particular, Eq.\ \eqref{errorss}, assures that $\bar{\chi}^{\text{ss}}_{z_0}(\bm{q},z)$ and $\bar{\chi}(\bm{q},z)$ are tangent to each other in $z_0$:
\begin{equation}
\frac{\partial\bar{\chi}^{\text{ss}}_{z_0}(\bm{q},z_0)}{\partial z} = \frac{\partial\bar{\chi}(\bm{q},z_0)}{\partial z}\label{tanchiss}
\end{equation}

\begin{figure}[h] 
    \centering
    \includegraphics[width = .9\columnwidth, keepaspectratio]{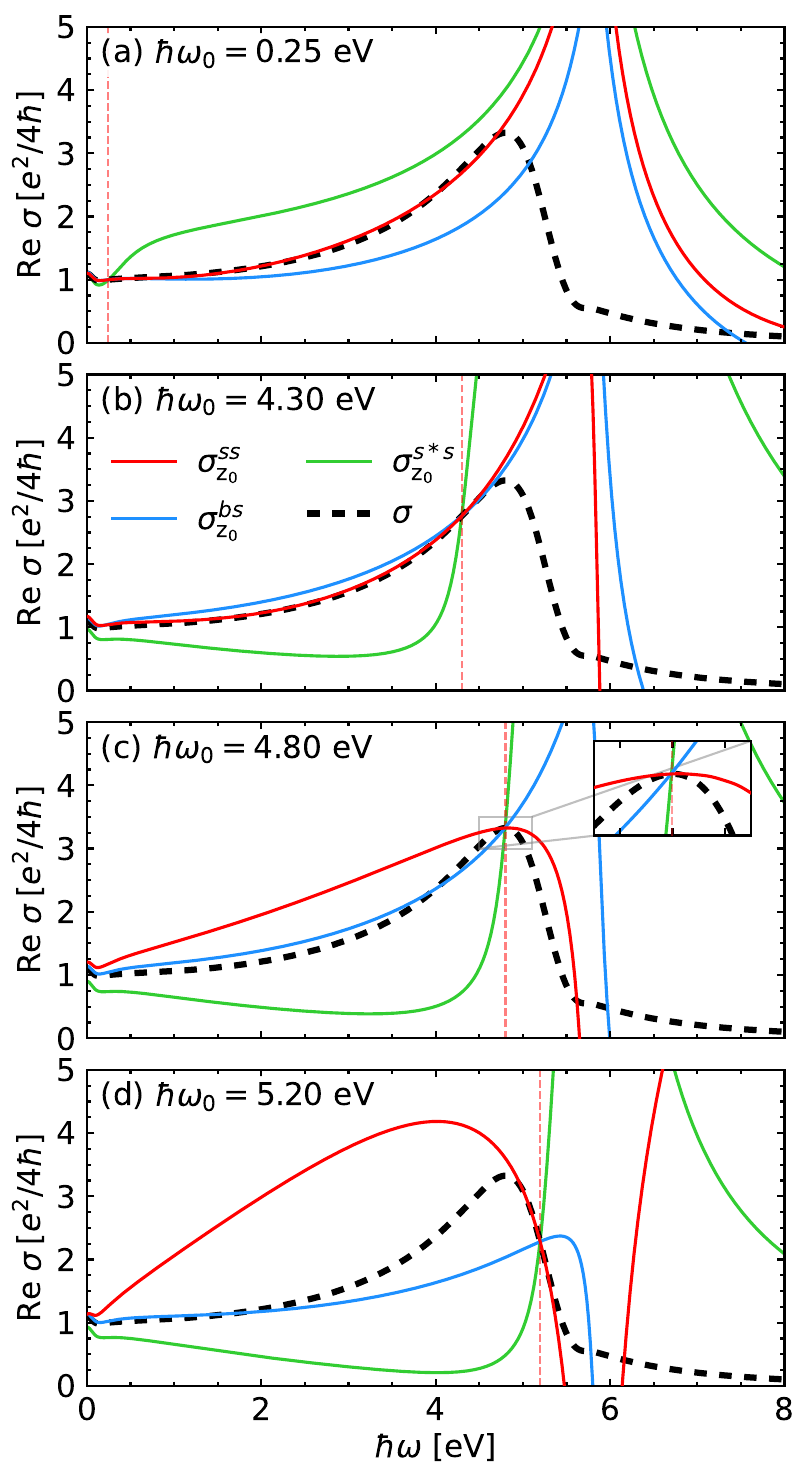}
    \caption{Fixed frequency approximation of the optical conductivity of graphene. The exact reference $\sigma(z)$ is indicated by black dashed line. The screen-screen, bare-screen, screen*-screen approximated conductivities $\sigma_{z_0}(z)$ (red, blue and green respectively) are obtained using Eqs.\ \eqref{chi sb in kspacez0}, \eqref{chi ss in k spacez0}, and \eqref{chi s*s in k spacez0}, where the screened vertexes are computed at a fixed reference value of $z_0{=}\omega_0{+}i\eta$. In each panel we indicate the $\hbar\omega_0$ in digits and
    by a vertical orange dashed lines. The screen-screen conductivity  is always tangent to the exact one in $\hbar\omega_0$, in contrast to the other approximated conductivities. Moreover, in all cases, but the one  of panel (c), where $\hbar\omega_0$ coincides with the peak, the screen-screen conductivity closely follows the  exact one on a wide range of energies. }
    \label{fig:sigma fixedw}
\end{figure}

Interestingly, by Eq.~\eqref{diffss}, we reformulate the response as that of an effective noninteracting electronic system (with, eventually, complex matrix elements). Such formulation does not require the expensive evaluation of the double-counting correction term, that involves a double summation (see second line of Eq.~\eqref{chi ss in k space}). Moreover, the fixed-frequency vertices are suitable for fast Wannier-function \cite{marini2024epiq,Lee2023} or other matrix-element interpolation schemes even in the present BSE approach. 
%%%riprendere in conclusioni (forse intro)
Finally, the presence of the differential operator $\Delta L^{0}$ localizes the contributions of interband transitions around the resonances with $z$ and $z_0$. This allows the restriction of the calculation to a limited number of bands near the Fermi-level, namely, to a low-energy model of the system.

We now discuss the numerical results. Let us call $\sigma^{\text{bs}}_{z_0}(z)$, $\sigma^{\text{ss}}_{z_0}(z)$ and $\sigma^{\text{s*s}}_{z_0}(z)$ the optical conductivity obtained with approximated vertices at $z{=}z_0$. Here, the aim is to compare numerically these three approximated expressions with respect to the exact $\sigma(z)$. Therefore, we select four representative frequency values $\hbar\w_0{=}\{0.25,4.3,4.80,5.20\}\, \text{eV}$, and represent the three different approximated conductivities in Fig.\ \ref{fig:sigma fixedw}(a)-(b)-(c)-(d) respectively. The numerical results confirm the validity of Eq.\ \eqref{tanchiss}. This is most evident in Fig.\ \ref{fig:sigma fixedw}(c) and (d). All the three formulations have the same value at the reference frequency, but only $\sigma^{\text{ss}}_{z_0}(z)$ is  tangent to the exact calculation. In Fig.\ \ref{fig:sigma fixedw}(c), $\hbar\w_0$ coincides with the maximum of reflectivity and the screen-screen approximation is in good agreement with the self-consistent calculation in frequency range of order ${\simeq}100\,\text{meV}$. In Fig.\ \ref{fig:sigma fixedw}(a)-(b)-(d) instead, $\sigma_{z_0}^{\text{ss}}(z)$ is in remarkable good agreement with $\sigma(z)$ for a much wider range of frequencies, of order of some eV, while $\sigma_{z_0}^{\text{bs}}(z)$ and $\sigma_{z_0}^{\text{s*s}}(z)$ perform much worse.
{ In these cases, we {attribute} the good agreement {between the full self-consistent calculation and the screen-screen approximation to} the fact that {$z_0$} is sufficiently far from both the Drude and the plasmon peak. Thus, the conductivity being rather flat, the screen-screen approximation works well since it posses exact first order derivative by construction.
Such consistency showcases the possibility of employing the screen-screen approximation to obtain precise calculations for a fraction of the computational cost. Moreover, the fixed frequency approximation could be used on an appropriate finite set of frequencies, to interpolate among different trends of the conductivity pertaining to different part of the optical spectrum. In this way, the fully self-consistent the optical conductivity can be reconstructed more precisely from few fixed-frequency approximations.}

\section{Conclusions}\label{sec: conclusions}
In the context of systems that can be described via a generalized Kohn-Sham equation, we have introduced three equivalent formulations for any electronic dynamical response function, referred to as bare-screen, screen-screen, and screen*-screen. We have shown that all the three rewritings of the response function can be expressed as values of appropriate functionals of the one-body electronic density matrix, but only the screen-screen response represents a stationary point of its functional. This implies that the error in the screen-screen response function is quadratic with respect to the error in the density matrix, while the bare-screen and screen*-screen formulations are instead linear. The variational character of the screen-screen response function is the central result of this work. It extends the findings of Ref.\ \cite{calandra2010adiabatic} obtained for nonadiabatic phonons within TD-DFT to any type of response function to incorporate the effect of nonlocal exchange-correlation effects. The screen*-screen formulation also holds considerable physical significance, as we show that its imaginary part can be interpreted as a generalized Fermi Golden Rule. 
Moreover, we have showed how the variational screen-screen formulation can be extended by including part of the interaction to the partially screened vertices and the rest to a partially interacting electron-hole propagator. The partial screen-partial screen reformulation of the response function so obtained can be advantageous to evaluate the BSE corrections to the phonon self-energy on top of a standard static phonon calculation, as we will show in Ref.\ \cite{usarticolo4}. 
Even if all our derivations were carried out neglecting memory effects in the electronic interaction, we have showed that such an assumption can be relaxed and our results can be generalized including memory effects in the Hartree-exchange correlation part of the kernel.

We validated our findings using a tight-binding model of graphene within the SX approximation of the electronic interaction. To provide numerical evidence of variationality, we studied the convergence trends of the optical conductivity $\sigma(\omega)$ within a self-consistent iterative procedure. The different formulations of the response function exhibit distinct convergence behaviors. In particular, the screen-screen response decays quadratically with the distance from the converged solution for the density matrix, while the bare-screen and the screen*-screen formulations converge only linearly. The variationality of the screen-screen response function can also be used to perform approximate calculations more efficiently. To demonstrate such capabilities, we have devised a fixed-frequency approximation of the density matrix and applied it to compute the optical conductivity of graphene. Such an approximation consists in determining the density matrix self-consistently at a specific frequency $\omega_0$, and then use it for all other frequencies, avoiding the costly self-consistent routine at all frequencies. Our results show that the screen-screen approximation is the closest to the full self-consistent calculation among the formulations considered, and it is the only one that matches the first order derivative of the optical conductivity at $\omega_0$, consistently with its variational nature. The quality of the approximation in a wider neighborhood of $\omega_0$ depends on the form of the function near the reference frequency: where the optical conductivity is quite flat, we find remarkable agreement in an interval of frequency of the order of several eV; where it is more peaked, the approximation is valid in a smaller neighborhood, of the order of ${\simeq}100$ meV. In contrast, the quality of the bare-screen and screen*-screen approximations rapidly deteriorates with the distance from $\omega_0$.

{The core value of the results obtained with the fixed frequency approximation reside in the fundamental proof of concept of the variationality of the screen-screen formulation. {This opens} the door to {a whole set of} approximations that could significantly reduce computational costs when applied in {first-principles approached}. For instance, it could be used to compute the response function using a density matrix approximated at a high electronic temperature or on a coarse $k$-grid \cite{calandra2010adiabatic}, {the size of the grid being} among the main bottlenecks for BSE approaches.}

{Most importantly, we showed that in the screen-screen fixed-frequency approximation, the response function are written as the response of an noninteracting system with effective couplings. Such rewriting of the response is localized in energy --- i.e.\ pertaining only few electronic bands--- and does not feature any detail of the interaction. As such, it could be extremely relevant to construct low-energy models, in which effective static couplings are used \cite{cappelluti2012charged,villani2024giant}.
}

\section*{Acknowledgements}
We acknowledge useful discussions with prof.\ Matteo Calandra and prof.\ Gianluca Stefanucci. G.C.\ acknowledges fruitful discussions with Dr.\ Guglielmo Marchese. We acknowledge the MORE-TEM ERC-SYN project, grant agreement no.\ 951215. We acknowledge the EuroHPC Joint Undertaking for awarding this project access to the EuroHPC supercomputer LUMI, hosted by CSC (Finland) and the LUMI consortium through a EuroHPC Regular Access call. 

\appendix

\section{Derivation of the induced density in time-dependent linear response} \label{sec-app: td linear response}
In this Appendix, we derive the expression of the induced density Eq.\ \eqref{n = LV0} in terms of the bare electron-hole propagator $L^0$ from linear response theory. At linear order in the external field, the time-dependent generalized Kohn-Sham equation [Eq.\ \eqref{schrodinger exact}] can be recast in a Liouville equation for the induced density matrix $\hat n\pert$ \cite{rocca2008turbo}
\begin{equation}
    \label{app: lioville n1}
    i\hbar \pdv{t} \hat n\pert(t) = [\hat H\unpert, \hat n\pert(t)] + [\hat V_{\scf}\pert(t), \hat n\unpert]
\end{equation}
where $\hat H\unpert$ is the Hamiltonian in absence of the external field 
and $\hat n\unpert$ reads
\begin{equation}
    \label{app: n0}
    \hat n\unpert = \sum_{i}f_i\unpert |\psi_i\unpert\rangle\langle\psi_i\unpert|
\end{equation}
By using the evolution in the interaction picture
\begin{equation}
    \hat{\tilde{O}}(t) = e^{\frac{i}{\hbar}\hat H\unpert t} O(t) e^{-\frac{i}{\hbar}\hat H\unpert t} ,
\end{equation}
Eq.\ \eqref{app: lioville n1} can be recast as 
\begin{equation} \label{app: liouville n1 interaction}
    i\hbar \pdv{t} \hat{\tilde{n}}\pert(t) =[\hat{\tilde{V}}_{\scf}\pert(t), \hat {\tilde{n}}\unpert],
\end{equation}
where $\hat n\unpert {=}\hat{\tilde{n}}\unpert$. We can evaluate the right-hand side of Eq.\ \eqref{app: liouville n1 interaction} using Eq.\ \eqref{app: n0} as
\begin{align}
    [\hat{\tilde{V}}_{\scf}\pert(t), \hat {\tilde{n}}\unpert] = \sum_{ij}e^{i\omega_{ij}t}(f\unpert_j {-} f\unpert_i) \nonumber \\ \times \langle\psi_i\unpert|\hat V_{\scf}\pert(t)|\psi\unpert_j\rangle |\psi_i\unpert\rangle\langle\psi_j\unpert| \label{app: [v,n0]}
\end{align}
where
$\w_{ij}=(\epsilon\unpert_i - \epsilon\unpert_j)/\hbar$.

We integrate Eq.\ \eqref{app: liouville n1 interaction} in time from $-\infty$ to $t$, considering that $\hat{{V}}_{\scf}\pert(t){=}0$, for $t{<}t_0$, where $t_0$ is the starting time of the perturbation, and we obtain 
\begin{equation}
    \label{n1 = int t}
    \hat n\pert (t) {=} \frac{1}{i\hbar}e^{-\frac{i}{\hbar}\hat H\unpert t}\!\! \int\limits^t_{-\infty}\!\!\dd{t\p}[\hat{\tilde{V}}\pert_{\scf}(t\p), \hat{\tilde{n}}\unpert] e^{\frac{i}{\hbar}\hat H\unpert t} .
\end{equation}
Using Eq.\ \eqref{app: [v,n0]} and taking the Fourier transform as in \eqref{fourier transform z}, we have from \eqref{n1 = int t}
\begin{eqnarray}
    \hat n\pert(z) &=& \sum_{ij}\frac{f\unpert_j {-} f\unpert_i}{i\hbar}  |\psi_i\unpert\rangle\langle\psi_j\unpert|\int\limits^{\infty}_{-\infty} \dd{t}e^{i(z{-}\w_{ij})t}\nonumber\\ 
    & &\times \int\limits^t_{-\infty}\!\!\dd{t\p}e^{i\omega_{ij}t\p} \langle\psi_i\unpert|\hat V_{\scf}\pert(t\p)|\psi\unpert_j\rangle. \label{app: n(z) dt dt'}
\end{eqnarray}
The integral in $t$ in Eq.\ \eqref{app: n(z) dt dt'} can be solved integrating by parts, obtaining:
\begin{eqnarray}
    \hat n\pert(z) &=& \sum_{ij}\frac{f\unpert_j {-} f\unpert_i}{i\hbar}  |\psi_i\unpert\rangle\langle\psi_j\unpert|\int\limits^{\infty}_{-\infty} \dd{t}\nonumber \frac{ie^{i(z{-}\w_{ij})t}}{(z{-}\w_{ij})}\\ 
    &\times & e^{i\omega_{ij}t} \langle\psi_i\unpert|\hat V_{\scf}\pert(t)|\psi\unpert_j\rangle. 
    \label{app: n(z) dt}
\end{eqnarray}
In fact, the lower extremal value of the primitive is $0$ since $V_{\scf}(t){=}0$ for $t{<}t_0$, while the upper one is zero since for retarded quantities it holds $\Im z{>}0$. From Eq.\ \eqref{app: n(z) dt} it follows 
\begin{equation}\label{app: rho(z) = L psi i psi j}
    \hat n\pert(z) {=} \sum_{ij} \frac{(f\unpert_j {-} f\unpert_i)|\psi_i\unpert\rangle\langle\psi_j\unpert|\langle\psi_i\unpert|\hat V_{\scf}\pert(z)|\psi\unpert_j\rangle}{\hbar z {-} (\epsilon_i\unpert {-} \epsilon\unpert_j)} .
\end{equation}
Considering the matrix elements in Schrodinger representation of Eq.\ \eqref{app: rho(z) = L psi i psi j} with the identification $(1){\to}(b)$, we obtain  Eqs.\ \eqref{n(r) = L0V  (r)} and  \eqref{L0(r)}. 

\section{Details of numerical implementation on tight-binding model of graphene} \label{sec-app:TB implementation}
In this Appendix, we give some details about the numerical implementation of the density-density response functions described in Sec.\ \ref{sec: application to graphene}. We have employed a tight-binding model of graphene, thoroughly described in Ref.\ \cite{albertoTB}. 

By Fourier transforming $v$ and $W$ in Eq.~\eqref{kernel sx}, and exploiting the symmetries of the crystal, we evaluated the interaction kernel as 
\begin{widetext}
\begin{multline}
K^{n\bm k {+} \bm q, m\bm k}_{s\bm k' {+} \bm q, l\bm k'}=   \frac{2}{A}\sum\limits_{\bm G}
\langle\psi^{(0)}_{n\bm k+\bm q}|e^{i(\bm q+\bm G)\cdot \bm r}|\psi^{(0)}_{m\bm k}\rangle
    v^{\ZD}_{\bm G}(\bm q)
        \langle\psi^{(0)}_{l\bm k\p}|e^{-i(\bm q+\bm G)\cdot\bm r'}|\psi^{(0)}_{s\bm k\p+\bm q}\rangle\\
        -\frac{1}{A}\sum\limits_{\bm G\bm G'}
    \langle\psi^{(0)}_{n\bm k+\bm q}|e^{i(\bm k-\bm k\p+\bm G)\cdot\bm r}|\psi^{(0)}_{s\bm k\p+\bm q}\rangle
    W_{\bm G\bm G'}^{\ZD}(\bm k-\bm k\p)
    \langle\psi^{(0)}_{l\bm k\p}|e^{-i(\bm k-\bm k\p+\bm G')\cdot\bm r'}|\psi^{(0)}_{m\bm k}\rangle \ ,
\end{multline}
\end{widetext}
where $A$ is the area of graphene unit cell and $\lbrace \bm G \rbrace$ are graphene reciprocal lattice vectors. For graphene, the vectors $\bm q$, $\bm k$,  $\bm k'$, $\bm G$ lie in the $(x,y)$ plane. The matrix elements are calculated with a tight-binding model, while $v^{\ZD}$ and $W^{\ZD}$ are effective 2D bare and screened Coulomb interactions.
Within the rectangular orbital approximation, the $z$ dependence of the single particle states $\psi$ is approximated as
\begin{equation}
    \psi_{n\bm k}(\bm r_{\parallel},z) \approx \psi_{n\bm k}(\bm r_{\parallel})\frac{\Theta(|z|-d/2)}{\sqrt{d}} \ 
\end{equation}
where $\bm r_{\parallel}=(x,y)$, $d$ is graphene thickness and ${\Theta}$ is the Heaviside function. Such kind of approximations, where the orbital shape is simplified, are usually accurate for response functions where $|\bm q|d \ll 1$, as our case. 
Within the rectangular orbital approximation, the effective 2D bare interaction can be written as
\begin{align}
    v^{\ZD}_{\bm G}(\bm q) &= 
    \frac{1}{d^2}\int\limits_{-d/2}^{d/2} \dd{z} 
    \int\limits_{-d/2}^{d/2} \dd{z}'
    \int d(\bm r_{\parallel}-\bm r'_{\parallel}) \nonumber\\
    %v_{\bm G}(\bm q,z,z') 
   & \times\frac{e^{-i (\bm q + \bm G)\cdot (\bm r_{\parallel}-\bm r'_{\parallel})}}{\varepsilon_r\sqrt{(z-z')^2+(\bm r_{\parallel}-\bm r'_{\parallel})^2}} \nonumber
    \\
    &= \frac{2\pi}{\varepsilon_r|\bm q+\bm G|}F(|\bm q+\bm G|d)\label{eq:v_2D_fin_thick} \ ,
\end{align}
where $\varepsilon_r$ is the relative dielectric constant of an eventual uniform background and
\begin{equation}
    F(x) = 
    %\frac{1}{d^2} \int\limits_{-d/2}^{d/2} \dd{z} \int\limits_{-d/2}^{d/2} \dd{z}' e^{-|x/d||z-z'|}\\ = 
    \frac{2}{x}\left(1+\frac{e^{-x}-1}{x}\right) \ .
\end{equation}
The screened 2D effective Coulomb interaction can be written as
\begin{align}
    W^{\ZD}_{\bm G\bm G'}(\bm q) &=
    \frac{1}{d^2}\int\limits_{-d/2}^{d/2} \dd{z}
    \int\limits_{-d/2}^{d/2} \dd{z}'
    \int d\bm r_{\parallel}
    \int d\bm r'_{\parallel} \nonumber\\ &\times
    e^{-i(\bm q +\bm G)\cdot \bm r_{\parallel}}
    W(\bm r,\bm r')
    e^{i(\bm q +\bm G')\cdot \bm r'_{\parallel}} \nonumber
    \\
    &= \varepsilon^{-1,\ZD}_{\bm G\bm G'}(\bm q)v^{\ZD}_{\bm G'}(\bm q) \ ,
\end{align}
where $\varepsilon^{-1,\ZD}_{\bm G\bm G'}(\bm q)$ is the effective 2D static inverse dielectric constant
\begin{multline}
    \varepsilon^{-1,\ZD}_{\bm G\bm G'}(\bm q) =
    {\frac{1}{d^2}}\int\limits_{-d/2}^{d/2} \dd{z}
    \int\limits_{-d/2}^{d/2}\dd{z}'
    \int \dd{\bm r_{\parallel}}
    \int \dd{\bm r'_{\parallel}}\\
    \times e^{-i(\bm q +\bm G)\cdot \bm r_{\parallel}}
    \varepsilon^{-1}(\bm r,\bm r')
    e^{i(\bm q +\bm G')\cdot \bm r'_{\parallel}} \ .
\end{multline}

The procedure to obtain the dielectric  function $\varepsilon^{-1,\ZD}_{\bm G\bm G'}(\bm q)$, within the random-phase approximation and self-consistently with the SX band structure is detailed in Ref.\ \cite{albertoTB}. As in Ref.\ \cite{albertoTB}, we suppose that the screened interaction is well approximated by that of an uniform system, i.e. $W^{\ZD}_{\bm G \bm G'}(\bm k- \bm k') \approx \delta_{\bm G \bm G'}W^{\ZD}(\bm k- \bm k'+\bm G)$. Finally, the density-density
response computed in the condition of null macroscopic
electric field is obtained using $ \bar K^{n\bm k {+} \bm q, m\bm k}_{s\bm k' {+} \bm q, l\bm k'}$, that is given by Eq.~\eqref{eq:v_2D_fin_thick}, where in the first summation (in the Hartree contribution) the $\bm G = \bm 0$ term is omitted.

In order to obtain the density-density response function $\bar\chi(\bm q,z)$ [Eq.\ \eqref{chi sb in kspace}], we need to compute the density matrix $n^{(\bm q)} (z)$ on a set of $\omega$ points at fixed $\eta$. In our implementation of a tight-binding model of graphene, we solve numerically the reciprocal space expression for the self-consistent Eqs.\ \eqref{app: sternheimer} and \eqref{app: V_scf = V + Kn}. The electron energies $\lbrace\epsilon^{(0)}_{n \bm k}\rbrace$ and wavefunctions $\lbrace \psi^{(0)}_{n\bm k} \rbrace$ are obtained in a tight-binding model with parameters fitted to DFT ab-initio calculations, then employing many-body perturbation theory within the SX approximation to include nonlocal exchange effects.
The band indices $\{n,m,l,s\}$ in Eqs.\ \eqref{app: sternheimer}, \eqref{app: V_scf = V + Kn} and \eqref{kernel sx} indicate either $\pi/\pi^*$ bands.
In the calculation of the kernel, we selected the $\bm G$ vectors satisfying the condition $|\bm G|^2/2 < 2.5$ Ry, that corresponds to include $\bm G = (0, 0)$ and the first shell of the six $\bm G$ vectors with lower modulus.
We used also a cutoff in the screened interaction $W^{\ZD}(\bm k- \bm k'+\bm G)$, such that $W^{\ZD}(\bm k - \bm k' + \bm G) = 0$ if $|\bm k - \bm k' + \bm G| < |\mathrm{\boldsymbol{K}}|$, where $\mathrm{\boldsymbol{K}}$ is the Dirac point.
We have performed our calculations reported in Sec.\ \ref{sec: application to graphene} using a $361{\times}361$ Monkhorst-Pack grid in reciprocal space, and electronic smearing of $\eta{=}100\,\text{meV}$. The optical conductivities are obtained for a small but finite wavevector, of value $\bm q = (10^{-3}, 0) 2\pi/a$ in Cartesian coordinates, where $a=2.46\textup{~\AA}$ is the lattice parameter of graphene, and the $x$ axis is along the $\Gamma{\rm K}$ direction. We verified numerically that the same converge trends are shown at larger momentum transfer, at $\bm q = (0.1, 0) 2\pi/a$. For the convergence study presented in Sec.\ \ref{subsubsec: convergence study}, the reference conductivity $\sigma(z)$ is taken as $\sigma^{\text{ss}}_{{\tau}_\infty}(z)$, where $\tau_\infty {=} 41$ is the number of steps needed to achieve self-consistency in the screen-screen calculation.

{In Ref.\ \cite{albertoTB}, we have presented the numerical details of the solution of the BSE with a pseudo-Hermitian Lanczos algorithm. 
{In both the implentation we presented in Ref.\ \cite{albertoTB} and the one presented here, an iterative routine is employed to calculate the response function. The cycle is stopped when the relative error on the response function [in this case, the optical conductivity $\sigma_\tau(z)$] between two consecutive steps is below a certain convergence threshold. 
The pseudo-Hermitian Lanczos algorithm has the advantage that the full $\omega$ spectrum can be computed with just one iterative cycle.} {However, the self-consistent procedure presented here is more accurate for a single frequency calculation  with respect to the one we used in Ref.\ \cite{albertoTB}.
In fact, we tested that with the same parameters (k grid and $\eta$), we reached a converge threshold of $10^{-16}$ on the relative error of the response  with our procedure, while with the Lanczos algorithm we are not able to increase over a convergence threshold of $10^{-5}{-}10^{-8}$ (depending on the calculations parameters) due to numerical instabilities. Hence, the procedure devised in this work can be beneficial when the response function at a single frequency is needed, as it is the case for the calculation of phonon self-energy from the dynamical interatomic force constant matrix \cite{usarticolo4}. Moreover, even if implemented on a simple tight-binding model with only two bands ($\pi/\pi^*$), the self-consistent routine presented here is fitted for a more sophisticated implementation where empty states are not explicitly considered. In fact, resembling the TD-DFT routine, Eqs.\ (106) can be rephrased in terms of projectors into occupied states as per the Sternheimer equations, thus obtaining the set of self-consistent equations adopted in ab-initio implementations \cite{baroni2001phonons}.}}

We also note that both our self-consistent routine and the pseudo Hermitian Lanczos algorithm scale as $N_k^4$, where $N_k$ is the linear dimension of the k grid, while the exact diagonalization of the electron-hole Hamiltonian in the BSE  scales as $N_k^6$ (see Ref. \cite{albertoTB} for a formal definition of the BSE Hamiltonian).

\bibliography{variationalbib}

\end{document}